\documentclass{aa}
\usepackage{natbib}
\usepackage{graphicx}
\usepackage{amssymb}
\usepackage{color}
\usepackage{ulem}
\usepackage{booktabs}
\usepackage{amsmath}
\usepackage{stmaryrd}

\begin{document}

\title{Knotty protostellar jets as a signature of episodic protostellar accretion?}

\author{Eduard I. Vorobyov\inst{1,2,3},Vardan G. Elbakyan\inst{2}, Adele L. Plunkett\inst{4}, Michael
M. Dunham\inst{5}, Marc Audard\inst{6}, Manuel Guedel\inst{3}, and Odysseas Dionatos\inst{3}}
% List of institutions
\institute{ Institute of Fluid Mechanics and Heat Transfer, TU Wien, Vienna, 1060, Austria 
\and
Research Institute of Physics, Southern Federal University, Roston-on-Don, 344090 Russia 
\and
University of Vienna, Department of Astrophysics, Vienna, 1180, Austria 
\and
European Southern Observatory, Av. Alonso de Cordova 3107, Vitacura, Santiago de Chile, Chile
\and
Department of Physics, State University of New York at Fredonia, 280 Central Ave, Fredonia, NY 14063
\and
Department of Astronomy, University of Geneva, Ch. d'Ecogia 16, 1290 Versoix, Switzerland
}

\abstract 
{}
{We aim at studying the causal link between the knotty jet structure in CARMA~7, a young Class~0 protostar in the Serpens South cluster,  and episodic accretion in young protostellar disks.}
{We used numerical hydrodynamics simulations to derive the 
protostellar accretion history in gravitationally unstable disks around solar-mass protostars. We 
compared the time spacing between luminosity bursts $\Delta\tau_{\rm mod}$, caused by dense clumps spiralling on the protostar, with the differences of dynamical timescales between the knots  
$\Delta\tau_{\rm obs}$ in CARMA~7.}
{We found that the time spacing between the bursts have a bi-modal distribution
caused by isolated and clustered luminosity bursts. The former 
are characterized by long quiescent periods between the bursts 
with $\Delta\tau_{\rm mod} =\mathrm{a~few}~\times (10^{3}-10^4)$~yr,
whereas the latter occur in small groups with time spacing between the bursts 
$\Delta\tau_{\rm mod}= \mathrm{a~few}~\times (10-10^2)$~yr.
For the clustered bursts, the distribution of $\Delta\tau_{\rm mod}$ in our models 
can be fit reasonably well to the distribution of $\Delta\tau_{\rm obs}$ in the protostellar 
jet of CARMA~7, if a certain correction for the (yet unknown) inclination angle with respect to the
line of sight is applied.
The K-S test on the model and observational data sets suggests the best-fit values for 
the inclination angles of $55^\circ-80^\circ$, which become narrower ($75^\circ-80^\circ$)
if only strong luminosity bursts are considered.
The dynamical timescales of the knots in the jet of CARMA~7 are 
too short for a meaningful comparison with the long time spacings between isolated bursts in our models.
Moreover, the exact sequences of time spacings between the luminosity
bursts in our models and knots in the jet of CARMA~7 were found difficult to match.}
{Given the short time passed since the presumed luminosity bursts (from tens to hundreds years), 
a possible overabundance of the gas-phase CO in the envelope of CARMA~7 
as compared to what could be expected from the current luminosity may be used to confirm the burst nature
of this object. More sophisticated numerical models and observational
data on jets with longer dynamical timescales are needed to further explore the possible causal link between luminosity bursts and knotty jets.}

\authorrunning{Vorobyov et al.}
\titlerunning{Knotty jets and episodic accretion}

\maketitle

\section{Introduction}

Low-mass stars are generally understood to form due to the gravitational
collapse of dense gas and dust clouds known as prestellar cores. 
The bulk of the final stellar mass is accumulated in the early evolutionary phase 
known as the embedded phase of star formation when the nascent star 
is still surrounded by the infalling parental core. However, the manner in which the  
stellar mass is accumulated, via steady or time-varying accretion,  is poorly understood.  
A wide range of protostellar accretion rates inferred in the embedded phase 
\citep[e.g.][]{Enoch2009} suggests that accretion is time-variable, but other explanations 
are also possible.

In the simplest model of low-mass star formation, an isothermal
sphere collapses starting from the center of the core with the
mass infall rate $\dot{M}_{\rm infall} \sim c_{\rm s}^3/G$ \citep{Larson1969,Shu1977}, which
tapers off with time due to a depleting mass reservoir in the parental core 
\citep{VB2005}. 
Variations in the initial positive density perturbations and  
infall rates declining with time can in principle explain a wide range of accretion rates inferred
for young embedded stars without the need to invoke strong time variability.%\citep{Megeath}.

There is one caveat to this simple picture: the instantaneous infall rate $\dot{M}_{\rm infall}$ 
is not identical to the accretion rate on the star  $\dot{M}$, because most of the cloud 
core material  lands on a circumstellar disk before reaching the star. Various physical processes
that take place in the disk, such as the magnetorotational, thermal and gravitational 
instabilities, can trigger strong bursts  when the matter is transported through the disk towards the star 
\citep[e.g.][]{BL1994,Armitage2001,Zhu2009,VB2005,Dangelo2012,VB2015,Armitage2016,Meyer2017}. 
The prototypical examples of these bursts
are known as FU-Orionis-type (FUor) and EX Lupi-type (EXors) luminosity outbursts featuring an increase
in luminosity by 3-5 magnitudes compared to the pre-burst state \citep{Audard2014}, though the driving
force of these two types of outburst can be distinct. 
About half of known FUors are young embedded 
objects as indicated by silicate features in absorption \citep{Quanz2007} and the youngest 
known FUor, HOPS~383, belongs to the Class~0 phase \citep{Safron2015}.

Observational manifestations of accretion variability are not limited to energetic FUor and 
EXor bursts. Recent variability monitoring campaigns
\citep[e.g.][]{Pena2017,Herczeg2017} indicate that young embedded stars demonstrate intrinsic (not extinction
related) variability of different strengths and periods, ranging from days to years and showing 
both rises and dips in the light curves. Numerical models featuring gravitationally unstable disks
demonstrate accretion variability that can explain the observed range of mass accretion rates
in embedded star forming regions \citep{Vorobyov2009}, although these underestimate somewhat the 
amplitude of short-term variability on the order of months and  years
\citep{Elbakyan2016}. Models with gravitationally unstable disks
are also successful in resolving the luminosity problem \citep{Dunham2012}, according to which the mean
luminosity of embedded protostars is about an order of magnitude lower than
that predicted by the simple spherical collapse models \citep{Kenyon1990}.

Clearly, observations of FUors and variability monitoring campaigns can provide 
information on accretion variability on human-life timescales. But what about indicators 
of past variability?  Bursts are short-lived phenomena and may be simply missed. Accretion 
variability in general may be a transient
phenomenon, having periods of strong activity, alternated with longer quiescent phases. 
Fortunately, certain chemical species in the collapsing envelopes, such as CO, can retain 
signatures of past accretion bursts \citep{Lee2007,Visser2012,Vorobyov2013,Frimann2017,Rab2017}. 
Because typical freeze-out times of these species onto dust grains in the envelope (a
few kyr) are much longer than the burst duration (a few tens to hundred years), they can linger
in the gas phase in the envelope long after the system has returned into a quiescent stage, and
their abnormally high abundance can  be used to infer the past accretion bursts.

%References to Grinin?

In this paper, we focus on another phenomenon -- protostellar jets --
that may be used to trace back the history of protostellar accretion. 
Jets were first observed as a sequence
of shock fronts or knots seen at optical wavelengths and
known as "Herbig-Haro (HH) objects" \citep[e.g.][]{Reipurth1995}. 
Shocked gas from the jets can survive for thousands of years and
often propagate for a few pc from their driving source powered by disk accretion 
\citep{Reipurth1997}. 
The origin of the knots can be attributed to a launching mechanism at the jet
base that is variable in time (e.g. Bonito et al., 2010). 
Arce et al. (2007) summarized evidence in favor
that episodic variation in the mass-loss rate can produce a
chain of knotty shocks and bow shocks along the jet axis.
Here, we test a hypothesis that the presence of a sequence of knots in protostellar
jets is related to a time-variable driving force caused by
episodic accretion of matter from accretion disk onto the central protostar \citep[e.g.][]{Plunkett2015}. The knots can track many episodes of accretion bursts, and not just the most recent/strongest, as it
may be the case for the chemical tracers. We analyze the characteristics
of luminosity bursts of different strengths obtained in hydrodynamical models of gravitationally unstable
disks and compare them with available observational data on the time spacing of the knots in protostellar
jets. Complemented with predictions from disk chemical models, jets may provide invaluable constraints on models of episodic accretion, holding records on the number and time of intense accretion 
bursts during protostellar evolution \citep{Belloche2016}. 

The paper is organized as follows. In Sect.~\ref{model} we provide a brief description of the 
numerical model. In Sect.~\ref{analysisAR} and \ref{analysisL}, we analyse the characteristics of 
the  model accretion and luminosity bursts, respectively. The comparison of time spacings
between the bursts and differences in dynamical timescales of the knots is performed in 
Sect.~\ref{comparisonJB}. 
Our main conclusions and the model limitations are summarized in Sect.~\ref{summary}.

\section{Description of the model}
\label{model}
In this section we provide a brief description of the hydrodynamical model
used in this paper to derive the protostellar accretion rates.
A detailed description of the model can be found in \citet{VB2015}. 
We started our numerical simulations from the gravitational collapse
of a prestellar core and continued into the embedded phase of stellar
evolution, during which the protostar and protostellar disk were formed.
Simulations were terminated at the end of the embedded phase 
when most of the core had accreted onto the star plus disk 
system.  The age of the protostar at this instance was 0.3 Myr.

To save computational time and avoid too small time-steps, we introduce a
sink cell at $r_{\rm s.c.}=6$~AU. We impose free outflow boundary conditions so that the matter is
allowed to flow out of the computational domain but is prevented from
flowing in. The sink cell is dynamically inactive; it contributes
only to the total gravitational potential and secures a smooth behaviour
of the gravity force down to the stellar surface.
To accelerate the computations, the equations of mass, momentum, and 
energy transport are solved in the thin-disk limit, the justification of which 
is provided in \citet{VB2010}. The following physical processes 
in the disk are considered: disk self-gravity, cooling due to dust
radiation from the disk surface, heating via stellar and background irradiation,
and turbulent viscosity using the $\alpha$-parameterization. The $\alpha$-parameter
is set to a constant value of $5\times10^{-3}$.
The hydrodynamics equations  are solved in polar coordinates on a
numerical grid with $512\times512$ grid zones. The solution
procedure is similar in methodology to the ZEUS code \citep{SN1992} and 
is described in detail in \citet{VB2010}.

%\subsection{Initial conditions}

Our numerical simulations start from a pre-stellar core with the radial profiles
of column density $\Sigma$ and angular velocity $\Omega$ 
described as follows:
\begin{eqnarray}
\Sigma(r) & = & {r_0 \Sigma_0 \over \sqrt{r^2+r_0^2}}\:, \\
\Omega(r) & = &2 \Omega_0 \left( {r_0\over r}\right)^2 \left[\sqrt{1+\left({r\over r_0}\right)^2
} -1\right],
\label{ic}
\end{eqnarray}
where $\Sigma_0$ and $\Omega_0$ are the gas surface density and angular velocity 
at the center of the core. These profiles have a small near-uniform
central region of size $r_0$ and then transition to an $r^{-1}$ profile;
they are representative of a wide class of observations and theoretical models
\citep{Andre1993,Dapp09}. The core is truncated at $r_{\rm out}$, which is also the
outer boundary of the active computational domain (the inner boundary is at 
$r_{\rm s.c.}=6$~AU). Different values of $r_{\rm out}$ are chosen to generate cores of different 
mass $M_{\rm core}$. The central angular velocity $\Omega_{0}$
is chosen so as to generate cores with ratios of rotational to gravitational
energy $\beta$ that is consistent with the values inferred for pre-stellar
cores by \citet{Caselli2002}. The initial gas temperature is set to 10~K 
throughout the core, which is also the background temperature of external irradiation.

\begin{table*}
\center
\caption{\label{tab:1}Model parameters}
\begin{tabular}{cccccccc}
 &  &  &  &  &  &  &     \tabularnewline
\hline 
\hline 
Model & $M_{\mathrm{core}}$ & $\beta$ & $T_{\mathrm{init}}$ & $\Omega_{0}$ & $r_{0}$ & $\Sigma_{0}$ & $r_{\mathrm{out}}$ \tabularnewline
 & {[}$M_{\odot}${]} & {[}\%{]} & {[}K{]} & {[}$\mathrm{km\,s^{-1}\,pc^{-1}}${]} & {[}AU{]} & {[}$\mathrm{g\,cm^{-2}}${]} & {[}pc{]} \tabularnewline
\hline 
1 & 1.5 & 0.88 & 10 & 1.0 & 3400 & $3.7\times10^{-2}$ & 0.1  \tabularnewline
2 & 1.1 & 0.88 & 10 & 1.4 & 2400 & $5.2\times10^{-1}$ & 0.07 \tabularnewline
%3 & 1.1 & 0.88 & 25 & 5.7 & 960 & $3.2\times10^{-1}$ & 0.028 \tabularnewline
\hline 
\end{tabular}
\center{ $M_{\mathrm{core}}$ is the initial core
mass, $\beta$ the ratio of rotational to gravitational energy, $T_{\mathrm{init}}$ the
initial gas temperature, $\Omega_{0}$ and $\Sigma_{0}$ the angular velocity
and gas surface density at the center of the core, $r_{0}$ the radius
of the central plateau in the initial core, and $R_{\mathrm{out}}$ the initial radius of the
core}
\end{table*}

\begin{figure*}
\begin{centering}
\includegraphics[scale=0.6]{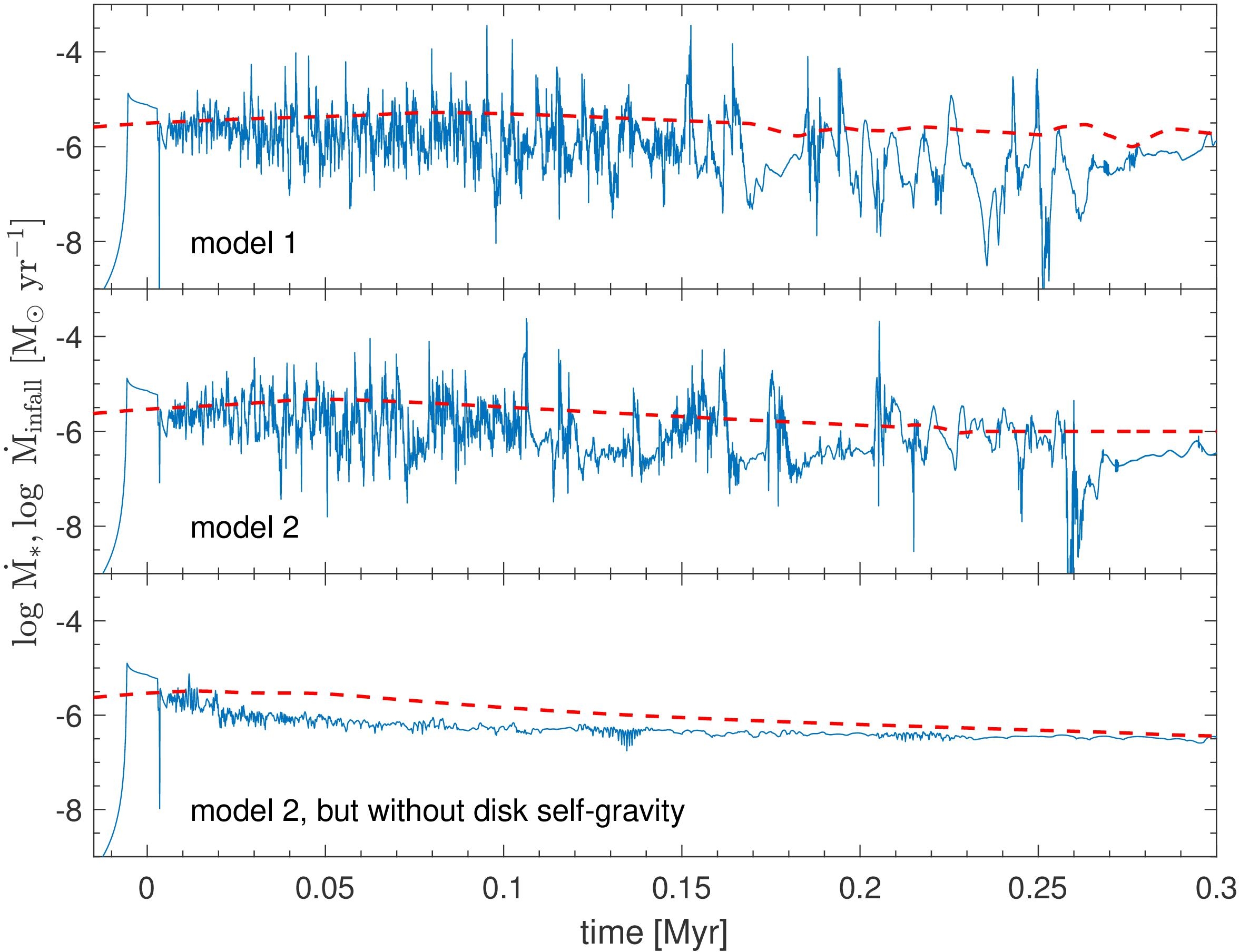} 
\par\end{centering}
\caption{\label{fig:1}Mass accretion rates on the central protostar (the blue solid
lines) and the disk infall rates (the red dashed lines) as a function of time in models~1 and 2.
The bottom panel shows the corresponding rates in model~2, but with disk self-gravity artificially turned
off.}
\end{figure*}

\section{Analysis of accretion and infall rates }
\label{analysisAR}
In this section, we analyze the mass transport rates defined as 
$-2\pi r \Sigma v_{\rm r}$, where $\Sigma$ is the gas surface density and 
$v_{\rm r}$ is the radial gas velocity. We  
calculated the mass transport rates through the sink cell at $r=6.0$~AU 
and at a distance of $r=2000$~AU from the central star. 
The former quantity serves as a proxy for the mass accretion rate onto the star $\dot{M}_\ast$ 
and the latter quantity
represents the mass infall rate onto the disk $\dot{M}_{\rm infall}$. We note that
the disk radius in our models is smaller than 1000~AU, but gravitational multi-body
scattering of gaseous clumps from the disk may produce perturbations in the gas flow in the vicinity
of the disk, so that we calculate $\dot{M}_{\rm infall}$ at a distance of 2000~AU
where the gas inward flow is unperturbed. We also note that
$\dot{M}_\ast$ may be modified by physical processes in the inner disk, such as 
the magnetorotational instability, not taken into account in our simulations. 
This may introduce additional variability to the protostellar
accretion rates obtained in our models and may explain why our models underestimate 
somewhat  the variability amplitudes at timescales on the order
of a few years and less when compared to observations \citep{Elbakyan2016}.

We consider two models, the parameters of which are presented in Table~\ref{tab:1}. 
The parameters of these models are similar to models 1 and 2 from \citet{VB2015}. 
Figure~\ref{fig:1} presents
$\dot{M}_\ast$ (blue solid lines) and $\dot{M}_{\mathrm{infall}}$(red
dashed lines) vs. time for our models. The evolutionary time presented in this work 
is counted from the instance of the central protostar formation rather than from the onset
of the gravitational collapse of pre-stellar cores because the duration of the 
collapse phase may vary from model to model.
Clearly, $\dot{M}_\ast$ demonstrates a time-variable behavior in both models,
whereas $\dot{M}_{\rm infall}$ is steady and gradually declines with time because of
gas depletion in the envelope. The mass infall
on the disk is one of the key parameters that triggers and sustains gravitational instability
and fragmentation in the disk \citep[e.g.][]{VB2005,Kratter2008} by replenishing the 
disk mass reservoir lost via accretion on the star.  A  decline of $\dot{M}_{\rm infall}$
results in weakening of both gravitational instability and accompanying accretion variability. 

As was previously shown in \citet{VB2010} and \citet{VB2015}, accretion variability is caused by 
a combination of two effects: the nonlinear interaction between different spiral
modes in the gravitationally unstable disk and infall of gaseous clumps formed in the disk via gravitational
fragmentation. In particular, the spiral arms produce regular, long-term variability
and the inspiralling clumps produce strong accretion bursts when destroyed and accreted by the star.
To demonstrate the influence of disk gravitational instability on the character of the protostellar
accretion rate, we show in the bottom panel of Figure~\ref{fig:1} the accretion and infall
rates obtained in model~2 but with disk self-gravity artificially turned off. Clearly, the
variability in $\dot{M}_\ast$ greatly reduced, whereas $\dot{M}_{\rm infall}$ remained
essentially similar to the case with disk self-gravity. It may appear as though 
the mean accretion rate is similar in both cases: with and without self-gravity. 
We checked the stellar masses $M_\ast$ at the same time instances and found that
the model with disk self-gravity has a systematically higher $M_\ast$. For instance,
$M_\ast=0.146~M_\odot$ in the model without disk self-gravity, whereas $M_\ast=0.315~M_\odot$
in the model with self-gravity. The mismatch becomes more pronounced with time. 
This means that the model with disk self-gravity is in fact characterized by
a systematically higher mean accretion rate, which is consistent with disk self-gravity
being the dominant mass transport mechanism in the early embedded phase of disk evolution
\citep{VB2009}.

To drive accretion variability on timescales of hundreds of thousands of years requires 
that large disk masses (or high disc-to-star mass ratios) be sustained for this length of time.
This can be achieved by continuing mass replenishment from the infalling envelope in the embedded
phase of star formation, lasting for about 0.1--0.5~Myr \citep[e.g.][]{Evans2009,Vorobyov2011}. High optical extinction and silicate features in
absorption towards about half of the known FUors imply that these objects are still embedded
in their parental cores \citep{Quanz2007,Audard2014} and, hence, may have massive disks.
Indeed, the recent estimates of disk
masses in three FUors \citep{Cieza2017} seem to confirm that they  possess
massive  disks ($M_{\rm disk}\ga 0.1 M_\odot$), sufficient
to trigger gravitational instability and fragmentation \citep[see fig. 1 in][]{Vorobyov2013}.

\section{Analysis of protostellar luminosities}
\label{analysisL}
In this section, we analyze the protostellar luminosities obtained in our models.
The total protostellar luminosity $L_{\mathrm{tot}}$ is calculated
as the sum of the photospheric luminosity $L_{\mathrm{ph}}$, arising
from the gravitational compression of the star and deuterium burning
in its core, and the accretion luminosity $L_{\mathrm{acc}}=G M_\ast \dot{M}_\ast/(2 R_\ast)$, 
arising due to the gravitational energy of the accreting matter. 
Here, $M_\ast$ and $R_\ast$ are the mass and radius of the protostar. The
former is calculated using $\dot{M}$ and the latter (and also $L_{\rm ph}$)
is computed using the Lyon stellar evolution code coupled to the main hydrodynamics code 
in real time as described in \citet{VB2015}.  

\begin{figure*}
\begin{centering}
\includegraphics[scale=0.6]{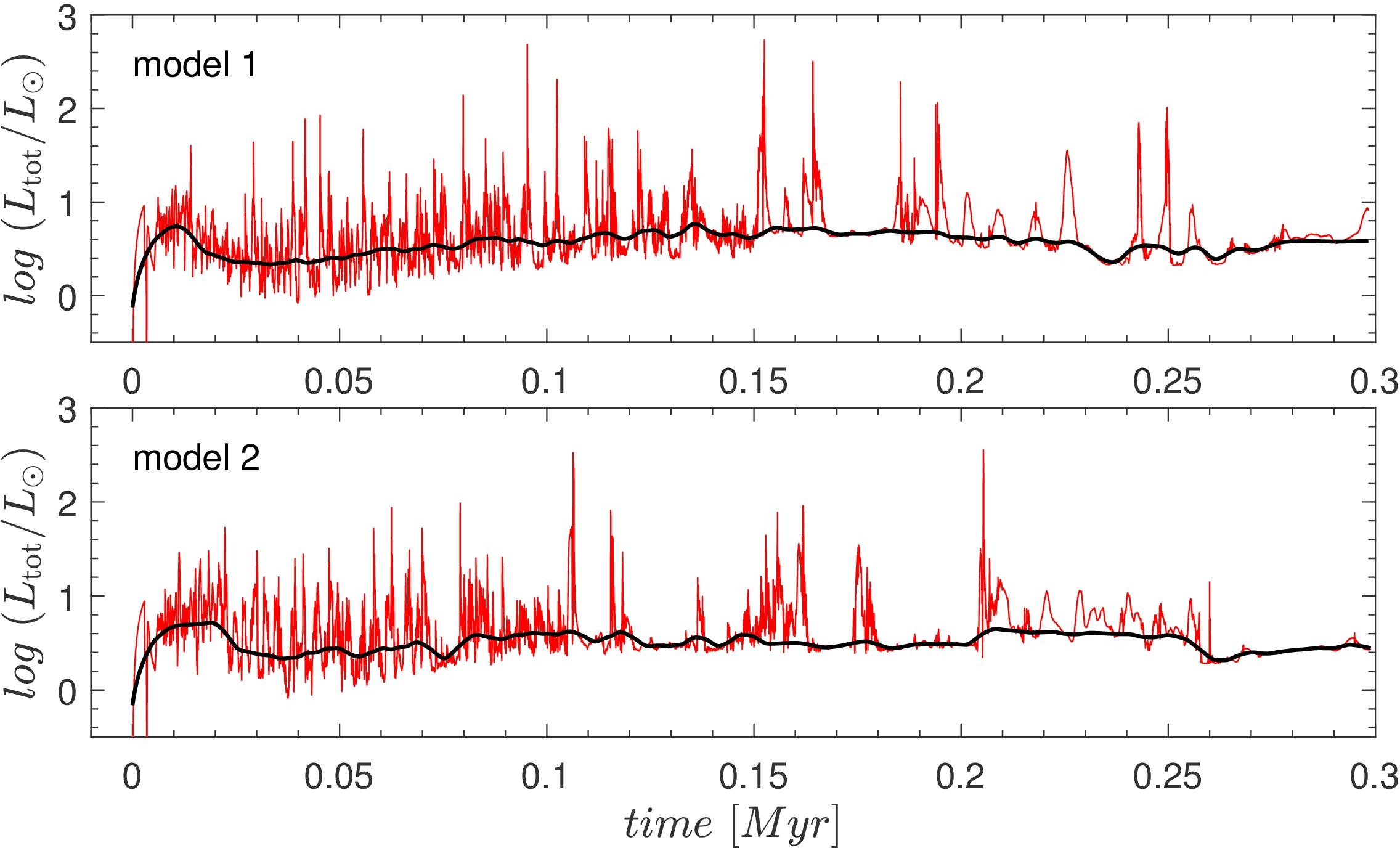}
\par\end{centering}
\caption{\label{fig:2} Total (accretion plus photospheric)
luminosity vs. time in models~1 and 2 shown by the red lines. 
The black solid line presents the background luminosity (see
text for more detail).}
\end{figure*}

The red lines in Figure~\ref{fig:2} present the  total luminosities as a function of
time for the same two models as in Figure~\ref{fig:1}. Clearly, $L_{\rm tot}$
demonstrates high variability with luminosity bursts of different strength, 
which is a direct consequence of variable protostellar
accretion rates. To investigate the possible causal link between episodic accretion and 
knotty jet structure, we need to distinguish short-lived luminosity bursts caused by infalling 
gaseous clumps from regular long-term variability caused by perturbations 
from large-scale spiral arms. To do this, we first introduce the background luminosity, in which
strong luminosity bursts were artificially filtered out as follows:
\begin{equation}
\begin{cases}
L_{\mathrm{bg}}(t)=\left\langle L_{\mathrm{ph}}(t)+L_{\mathrm{acc}}(t)\right\rangle , & \mathrm{if}\,\,\dot{M}_\ast(t)\leq\bar{\dot{M}}_\ast,\\
L_{\mathrm{bg}}(t)=\left\langle L_{\mathrm{ph}}(t)+(\bar{\dot{M}}_\ast/\dot{M}_\ast(t))L_{\mathrm{acc}}(t)\right\rangle , & \mathrm{if}\,\,\dot{M}_\ast(t)>\bar{\dot{M}}_\ast.
\end{cases}\label{eq:1}
\end{equation}
Here, $\dot{M}_\ast(t)$ is the instantaneous protostellar accretion rate and 
$\bar{\dot{M}}$ the accretion rate averaged over the age of the star in our simulations 
(which is 0.3~Myr). 
During the bursts, $\dot{M}_\ast(t)$ is much greater than $\bar{\dot{M}}$.
Multiplying $L_{\rm acc}(t)$ by $\bar{\dot{M}}_\ast/\dot{M}_\ast(t)$ means that 
the  instantaneous accretion rate $\dot{M}_\ast$ in the formula for $L_{\rm acc}$ 
is substituted with a time-averaged accretion rate $\bar{\dot{M}}_\ast$, which effectively
removes accretion bursts. In Equation~(\ref{eq:1}), the
angle brackets stand for time-averaging over a period
of $10^{4}$~yr with the LOcally WEighted Scatterplot Smoothing (LOWESS)
method using weighted linear least squares and a second-order polynomial
model \citep{Cleveland1998}.
The smooth background luminosities for each model are presented
in Figure~\ref{fig:2} by the black solid lines. 

As a second step
to distinguish luminosity bursts from regular variability, we 
make an assumption that $L_{\rm tot}$ must be at least 2.5 times higher (one magnitude in brightness)
than $L_{\rm bg}$ during the burst. The duration
of each burst must be less than 500 years to filter out slow rises and drops in luminosity.
The bursts that are  2.5, 6.25, 15.6, and 39 times (1, 2, 3, and 4 magnitudes) higher than $L_{\rm bg}$
are hence called 1-mag, 2-mag, 3-mag, and 4-mag bursts, respectively. In addition,
we require that $L_{\rm tot}$ on both sides of the peak value
drop at least by a factor of 2.5  to filter out small kinks in a smoothly increasing or decreasing 
luminosity curve. 

\begin{figure*}
\begin{centering}
\includegraphics[scale=0.8]{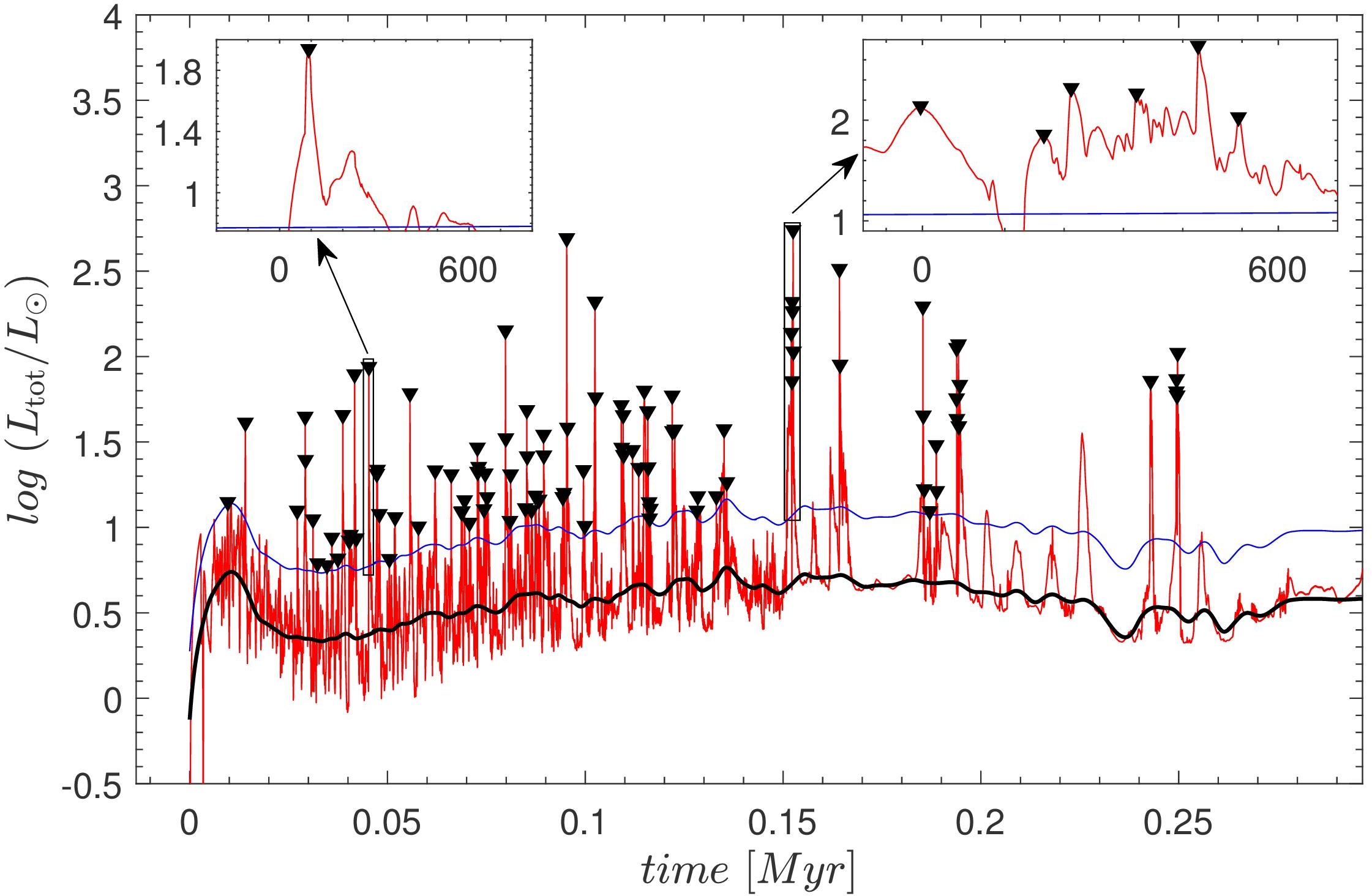} 
\par\end{centering}
\caption{\label{fig:3} Total luminosity (the red lines) and background luminosity (the thick 
black lines) vs. time in model~1. The blue line shows the 1-magnitude
cutoff above the background luminosity  and the black triangles mark the luminosity bursts. 
The right and left insets present examples of the isolated and clustered luminosity bursts. See text
for more detail.}
\end{figure*}

The resulting luminosity bursts in model~1
are marked in Figure~\ref{fig:3} with the filled triangles. The red and black lines show the
total and background luminosity, while the blue line represents the 1-mag 
cut-off used to distinguish the bursts from regular variability.
Clearly, model~1 shows a number of bursts of various magnitude. Some bursts are isolated while others
are closely packed. 
The insets in Figure~\ref{fig:3} zoom in on short time periods of evolution featuring 
an isolated luminosity burst (left) and a series of closely packed bursts (right). 
The former are caused by compact and dense clumps that withstand the tidal torques when approaching
the star until they are accreted through the sink cell, while the latter are caused by extended 
fragments that were stretched by tidal torques in a series of smaller clumps 
when approaching the star, with each of the smaller clumps causing one burst \citep[see fig. 13 in][]{VB2015}. 
V1057~Cyg is a prototype of isolated luminosity bursts, showing a steep rise followed by a slow decline
over timescales of several tens of years, while V346~Ori may present an observational example
of the clustered burst, the light curve of which shows an intermittent pattern with a deep 
drop between the previous and the current burst \citep{Kraus2016,Kospal2017}.

\begin{table*}
\center
\caption{Summary of burst characteristics}
\label{tab:2}
\begin{tabular}{ccccccc}
 &  &  &  &  &  & \tabularnewline
\midrule
\midrule 
 & Model  & $N_{\mathrm{bst}}$  & $L_{\mathrm{max}}/L_{\mathrm{min}}/L_{\mathrm{mean}}$  & $\dot{M}_{\mathrm{max}}/\dot{M}_{\mathrm{min}}/\dot{M}_{\mathrm{mean}}$  & $t_{\mathrm{bst}}^{\mathrm{max}}$/$t_{\mathrm{bst}}^{\mathrm{min}}$/$t_{\mathrm{bst}}^{\mathrm{mean}}$  & $t_{\mathrm{bst}}^{\mathrm{tot}}$\tabularnewline
 &  &  & {[}$L_\odot${]}  & {[}$\times10^{-5}\,M_\odot\mathrm{y}\mathrm{r}^{-1}${]}  & {[} yr {]}  & {[} yr {]}\tabularnewline
\midrule 
\multicolumn{7}{c}{1-mag cutoff}\tabularnewline
\midrule 
 & 1  & 48  & 30/6/14 & 1.8/0.36/0.95 & 241/11/102  & 4790\tabularnewline
& 2  & 46  & 30/7/13 & 1.4/0.32/0.80 & 243/9/100  & 4528\tabularnewline
%& 3  & 24  & 125/23/62 & 2.2/0.63/1.43 & 240/31/137  & 3149\tabularnewline
%& Total  & 118  & 125/6/30 & 2.2/0.32/1.06 & 243/9/113
%& 4156{*}\tabularnewline
\midrule 
\multicolumn{7}{c}{2-mag cutoff}\tabularnewline
\midrule 
 & 1  & 30  & 70/20/40 & 4.8/1.4/2.5  & 220/6/55  & 1586\tabularnewline
& 2  & 27  & 45/14/28  & 3.1/0.44/2.03 & 208/13/57  & 1494\tabularnewline
%& 3  & 15  & 310/134/199  & 6.5/2.07/4.02 & 99/8/57  & 791\tabularnewline
%& Total  & 72 & 310/14/89  & 6.5/0.44/2.85 & 220/6/56 & 1290{*}\tabularnewline
\midrule 
\multicolumn{7}{c}{3-mag cutoff}\tabularnewline
\midrule 
& 1  & 16 & 180/44/91  & 11.8/1.9/5.5 & 191/6/47  & 700\tabularnewline
 & 2  & 8 & 97/53/81  & 9.1/3.9/5.7 & 32/5/19  & 130\tabularnewline
%& 3  & 4 & 620/362/486  & 11.6/9.13/9.98 & 82/37/55  & 165\tabularnewline
%& Total  & 28 & 620/44/219 & 11.8/1.9/7.06 & 191/5/40 & 332{*}\tabularnewline
\midrule 
\multicolumn{7}{c}{4-mag cutoff}\tabularnewline
\midrule 
& 1  & 7  & 537/192/323  & 36.1/7.95/21.0  & 58/8/21  & 124\tabularnewline
& 2  & 6  & 357/197/256  & 20.6/11.5/16.8  & 17/7/13 & 63\tabularnewline
%& 3  & 0  & -  & -  & -  & -\tabularnewline
% & Total  & 13  & 537/192/290  & 36.1/7.95/18.9  & 58/7/17  & 94{*}\tabularnewline
\bottomrule
\end{tabular}
\center{$N_{\mathrm{bst}}$ is the
number of bursts of the corresponding magnitude, $L_{\mathrm{max}}/L_{\mathrm{min}}/L_{\mathrm{mean}}$
are the maximum, minimum and mean total luminosities, respectively, $\dot{M}_{\mathrm{max}}/\dot{M}_{\mathrm{min}}/\dot{M}_{\mathrm{mean}}$
are the maximum, minimum and mean accretion rates through
the central sink cell, $t_{\mathrm{bst}}^{\mathrm{max}}$/$t_{\mathrm{bst}}^{\mathrm{min}}$/$t_{\mathrm{bst}}^{\mathrm{mean}}$
are the maximum, minimum and mean durations of the bursts, $t_{\mathrm{bst}}^{\mathrm{tot}}$
is the total duration of all the burst with the corresponding magnitude.}
%The sign $^{*}$ denotes the mean of $t_{\mathrm{bst}}^{\mathrm{tot}}$
%for all models.}
\end{table*}

For each of 2 models listed in Table \ref{tab:1}, we calculated the number of 1-, 2-, 3- and 
4-mag bursts $N_{\mathrm{bst}}$. In particular, by the 1-mag burst we mean all bursts with luminosity
greater than 2.5 times the background luminosity (1-mag cutoff), 
but is lower than 2.5$^2$ times the background luminosity (2-mag cutoff). The 2-, 3-, and 4-mag 
bursts are defined accordingly, with an exception of the 4-mag bursts having no upper limit. 
The results are presented in the second column of 
Table~\ref{tab:2}. Clearly, the number of bursts rapidly decreases from 1- to 4-mag ones. 
This can be understood as a consequence of tidal stretching and disruption of massive extended 
fragments into smaller clumps when approaching the star, making strong bursts a less likely
outcome. The mass function of forming clumps is also skewed towards objects of smaller mass, from a
few Jupiter masses to a few tens of Jupiter masses \citep{Vor2013}. 

%One can see a notable 
%decrease in the number
%of bursts in model~3 as compared to the other two models. This is caused by a generally warmer disk
%in the $T_{\rm init}=25$~K model~3, making gravitational instability and fragmentation less vigorous.
%Accretion of some massive clumps on central star from the gravitationaly unstable
%protostellar disk, can lead to the increase of mass accretion rates
%up to the few$\times10^{-4}\,M_{\varodot}\mathrm{y}\mathrm{r}^{-1}$.
%This leads to steep rise of stellar luminosity from few solar up to
%tens, and even hundreds of solar luminosities, which is typical for
%FUOrs. 

The peak values of stellar luminosity and mass accretion rate during the bursts are presented 
in the third and forth column of Table \ref{tab:2}. In particular, the third column shows the 
maximum luminosity $L_{\mathrm{max}}$, minimum luminosity
$L_{\mathrm{min}}$ and arithmetic mean luminosity $L_{\mathrm{mean}}$ for all bursts in each
model. The fourth column shows the corresponding values for accretion
rates: $\dot{M}_{\mathrm{max}}$, $\dot{M}_{\mathrm{min}}$, and $\dot{M}_{\mathrm{mean}}$. 
As can be expected, the luminosities and mass accretion rates increase with increasing magnitude of
the burst. 

%One can also notice that model~3 with a higher initial (and background) temperature 
%features bursts with on average higher luminosities. There are two reasons for this phenomenon. 
%First, 
%the background luminosity is on average higher in this model due to a higher mass accretion rate
%(and associated accretion luminosity) caused by a higher mass infall rate from a warmer parental
%cloud ($\dot{M}_{\rm infall} \propto c_{\rm s}^3/G$). Second, the mass of clumps forming in the disk
%via gravitational fragmentation is also higher, causing bursts of stronger magnitude, because 
%the Jeans mass in a warmer disk of model~3 is higher.

\begin{figure}
\begin{centering}
\resizebox{\hsize}{!}{\includegraphics{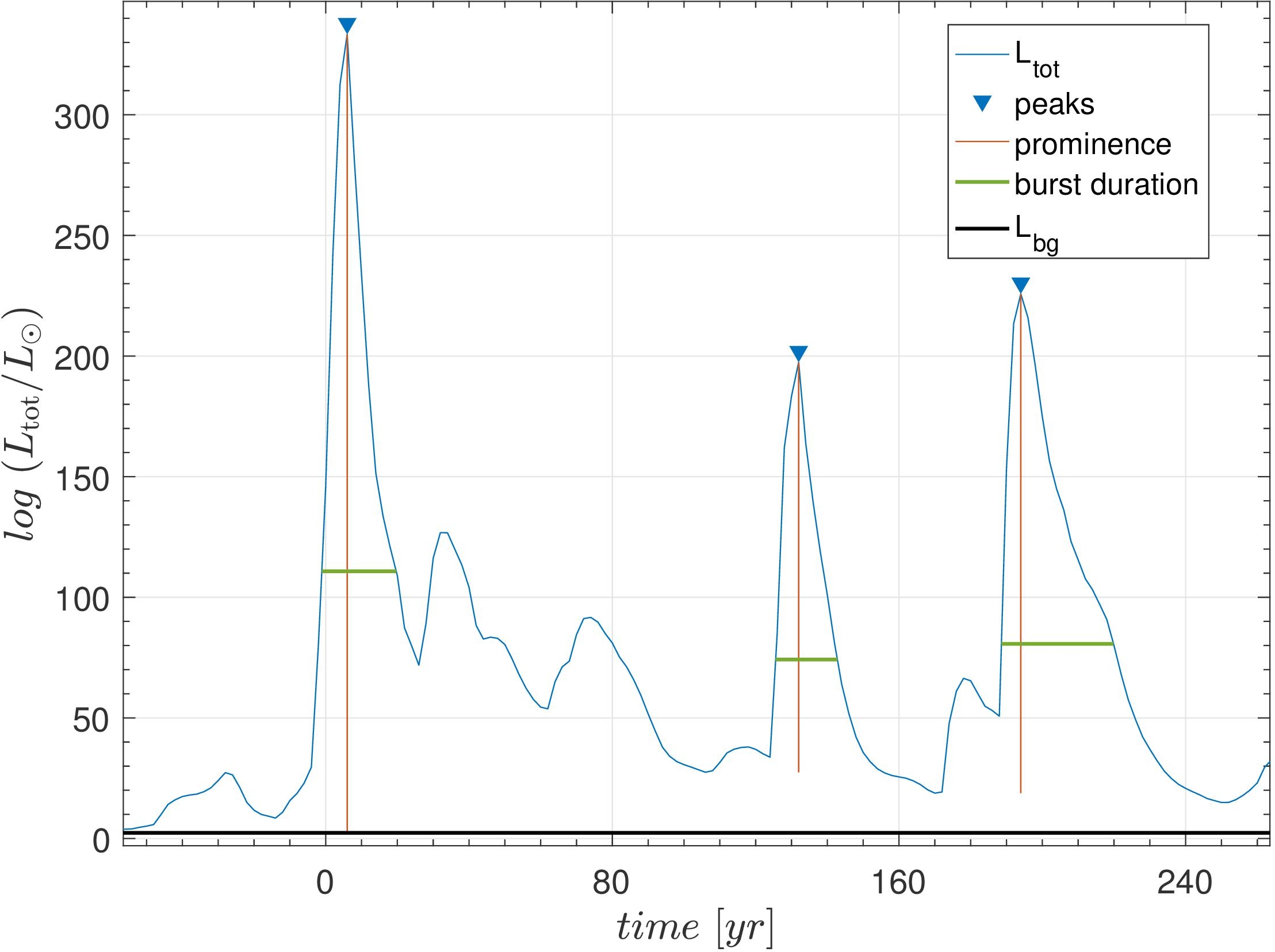}}
\par\end{centering}
\caption{\label{fig:9}Total luminosity vs. time plot (the blue line)) 
illustrating our procedure for calculating  the characteristics of the bursts 
including their prominence (the vertical red lines) and duration (the horizontal green lines). }
\end{figure}

To calculate the burst duration, we analyzed the shape of each burst as illustrated in Figure~\ref{fig:9}.
More specifically, we calculated the duration of each burst with respect to the
prominence of the burst shown by the red vertical lines. The prominence of the burst measures how
much the burst stands out due to its intrinsic height and its location
relative to other peaks. To calculate the prominence, we extend a horizontal line from the peak 
to the left and right until the line does one of the following: a) crosses the mass accretion curve
(the blue line) because there is a higher peak, b) reaches the left or right end of the
mass accretion curve. We then find the minimum of the mass accretion curve 
in each of the two intervals defined in the previous step. The higher of the two minima specifies 
the reference level. The height of the peak above this level is its prominence.
We assume the duration of the burst equal
to the distance between the points where the descending total luminosity (on both sides from the peak)
intercepts the green horizontal line beneath the peak at a vertical distance
equal to 1/3  of the burst prominence. Clearly, there is certain
freedom in choosing the position of the horizontal lines, but our resulting values of the burst durations
are in reasonable agreement with what is known about durations of FUor
luminosity outbursts \citep{Audard2014}.
The maximal duration ($t_{\mathrm{bst}}^{\mathrm{max}}$),
minimal duration ($t_{\mathrm{bst}}^{\mathrm{min}}$) and arithmetic mean of burst durations 
($t_{\mathrm{bst}}^{\mathrm{mean}}$) are presented in the fifth column of Table~\ref{tab:2}. 
Strong 3- and 4-mag luminosity bursts that are most relevant to FUor luminosity outbursts,
have burst durations ranging from several years to tens of years, with the longest duration of 191~yr.
These values are consistent with durations of FUor outbursts \citep{Audard2014}, especially if we take
into account that the longest outburst, that of FU Orionis itself, has been in the 
active state for already more than 80 years and so far shows no sign of fading.
The durations of weaker 1- and 2-mag bursts are a factor of several longer (because
they are caused by accretion of tidally stretched clumps), though remaining 
within reasonable limits. 
We note that such bursts may be more difficult to observe, especially in the
deeply embedded phases of stellar evolutions. We also calculated the total duration of bursts 
($t_{\mathrm{bst}}^{\mathrm{tot}}$) for each model. The resulting values are presented in the 
sixth column of Table~\ref{tab:2}.  Clearly, the total duration of the bursts is much shorter 
than the considered evolution
time of a few hundred thousand years. This explains why luminosity bursts are rarely observed, 
but must be numerous during the early evolution of young protostars.

In our models, the mass accretion rate $\dot{M}_\ast$ is calculated at the
position of the inner sink cell, $r_{s.c.} = 6$~AU.  The question that arises
is how much $\dot{M}_\ast$ can be sensitive to the choice of  $r_{s.c.}$.
We varied the value of $r_{s.c.}$ in other studies from 10~AU \citep{VB2005,VB2015} 
to 2.0 AU \citep{Elbakyan2016} and found little difference in the qualitative 
behaviour of the mass accretion rate. On the other hand, as fragments formed in the outer 
disk approach the star,
they must be inevitably stretched out due to tidal torques. How
much of the fragment material finally reaches the star and how much is retained by the inner disk 
is an open question \citep[e.g.][]{NL2012}. An additional complication
is that the very inner disk regions ($\la$~a few AU) may trigger accretion bursts on their own. 
The effect of a sudden mass deposition onto the inner disk, as if
by infall of a fragment migrating through the disk onto the star,
has been investigated by \citet{Ohtani2014}. It was
found that such an event can lead to the FU-Orionis-like
eruption due to triggering of the magneto-rotational instability
 at sub-AU scales. The effect of the inner disk on the mass accretion rate history
requires focused high-resolution studies, which are planned for the near future.

\section{Knotty jets and episodic bursts}
\label{comparisonJB}
Jets from young stellar objects have been known for over three decades. 
Despite the fact that
the mechanisms responsible for launching the protostellar jets are not fully understood, it is 
generally accepted that there exists a causal link between the jet launching process and dynamical interaction
of accreted matter with the stellar and/or disk magnetic field \citep{Frank2014}. 
Historically, jets were first observed as a sequence of knots seen at optical wavelengths and 
known as "Herbig-Haro (HH)  objects", and now many HH objects are known  
\citep[e.g.][]{Reipurth1995}.  
The swept up molecular gas is known as the molecular outflow, and thought to be another 
manifestation of the same mass loss process from the forming star plus disk system.  
Molecular jets can also show knot morphology.  The origin of these knots can be attributed to 
a launching mechanism at the jet base that is variable in time \citep[e.g.][]{Bonito2010}. 
%The knotty structure seen in many protostellar 
%jets and spacing between the knots are consistent
%with short periods of variability in the mass accretion rates as could
%be expected from clustered bursts (Vorobyov\&Basu 2015). 
For instance, \citet{Dopita1978} and  \citet{Reipurth1997} suggested that short period 
intense accretion events caused by instabilities in the FU Orionis-type accretion disks 
could be responsible for the Herbig-Haro flows with multiple working surfaces.
\citet{Arce2007} summarized evidence in favor that episodic variation in
the mass-loss rate can produce a chain of knotty shocks and bow shocks
along the jet axis.  Observations of jets on spatial scales
of hundreds to thousands of AU make it possible to investigate mass accretion variability on dynamical
timescales up to a few thousand years, much longer than what is possible
with a series of direct observations of the accretion process \citep{Ellerbroek2014}.

Observations of knots at different
epochs can reveal how the outflow varies with time. This can help to reconstruct
the time periods when the knots were formed and reveal their pattern of occurrence. 
In most cases, however, observations of the line-of-sight velocities
from one epoch are only available. In this paper, we make use of the data on CARMA~7, a Class~0 object in the Serpens South cluster  observed by \citet{Plunkett2015} using the Atacama
Large Millimeter/sub-millimeter Array (ALMA). The outflow ejecta revealed 
22 knots (11 in each direction), the most recent having the highest line-of-sight velocity. 
The dynamic timescale for each knot was estimated as $\tau_{\mathrm{obs}}=D/V_{\mathrm{flow}}
(\mathrm{cos}\,i/\mathrm{sin}\,i)$,
where $D$ is the distance (projected on the plane of the sky) between the knot and protostar, $V_{\mathrm{flow}}$
is the line-of-sight velocity of the knots, and $i$ is
the (unknown) inclination of the outflow with respect to the line of sight.
The authors found dynamic timescales for each of the identified knots
ranging from 100 years to 6,000 years (uncorrected for the inclination
angle, discussed later in this section), assuming that the knots travel with constant velocity from
the time of their launch. They also made a suggestion that knots
might be related to an episodic ejection mechanism, such as accretion
bursts caused by disk instabilities. In this case, the difference between the dynamic
timescales of the knots $\Delta\tau\mathrm{_{obs}}$ can provide the durations
of quiescent phases between episodic ejections.

\begin{figure}
\begin{centering}
\includegraphics[width=1\columnwidth]{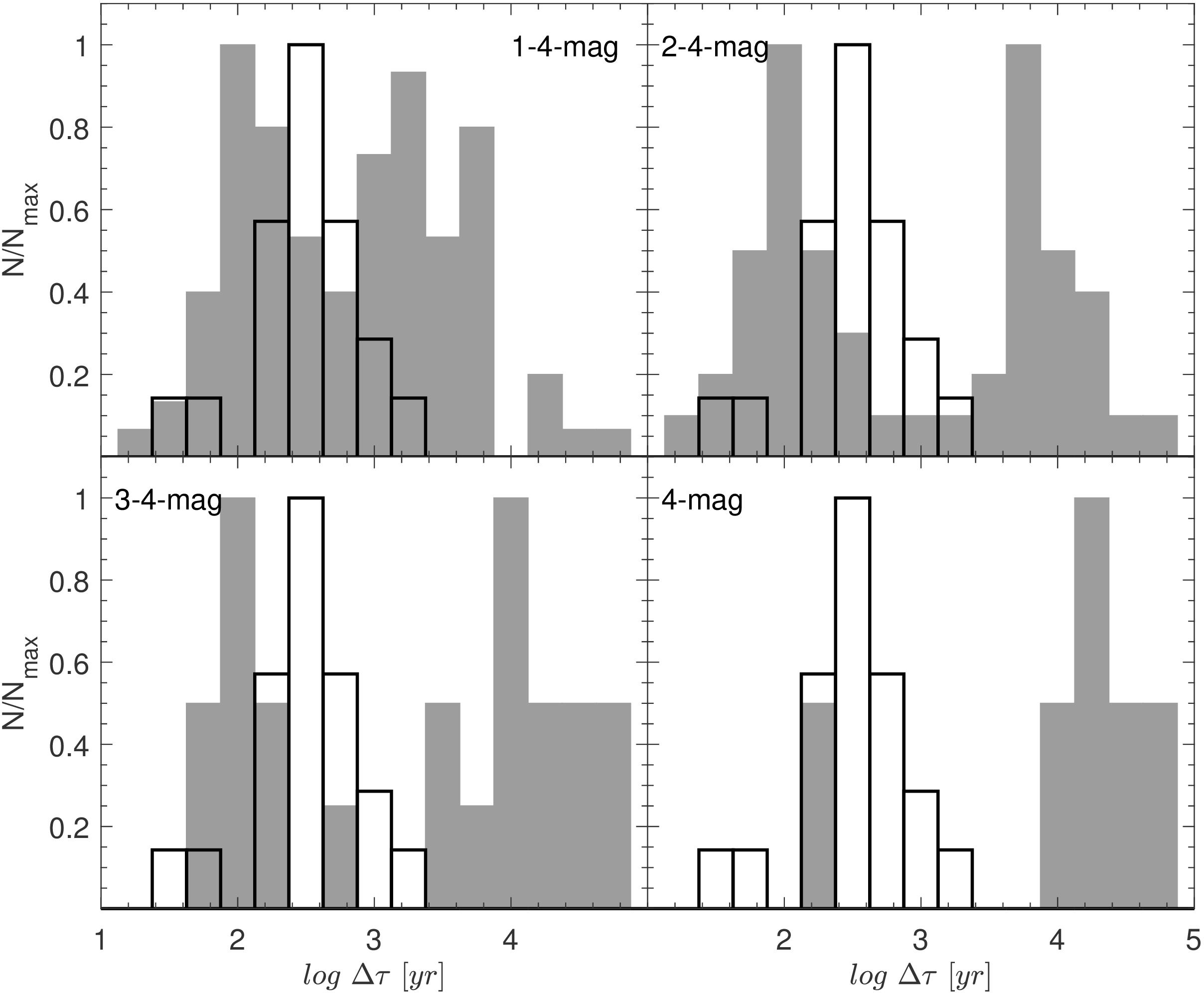} 
\par\end{centering}
\caption{\label{fig:4}Normalized distribution of time spacings between the luminosity bursts ($\Delta\tau_{\rm mod}$) of various magnitudes (1-4-mag, 2-4-mag, 3-4-mag, and 4-mag) 
in model~1 shown with the filled histogram.
The solid line histogram shows the distribution of differences in dynamical timescales
between the knots ($\Delta\tau_{\mathrm{obs}}$) in CARMA~7 without correction
for an inclination angle.}
\end{figure}

We can now calculate the time spacings $\Delta\tau_{\rm mod}$ between the 
subsequent bursts in our models and 
compare them with the differences in dynamic timescales between the knots $\Delta\tau_{\rm obs}$ 
in the jet of CARMA~7. More specifically, we calculate $\Delta\tau_{\rm mod}$ as the time spacing
between two consequent luminosity peaks comprising a range of magnitudes. For instance, 
$\Delta\tau_{\rm mod}$ for luminosity bursts of 1-4-mag signifies a time spacing between two consequent
luminosity peaks of all magnitudes, whereas $\Delta\tau_{\rm mod}$ for luminosity bursts
of 4-mag denotes a time spacing between two luminosity peaks of 4-mag (skipping all bursts of lesser
magnitude).
%As a next step, to make a comparison with observations of episodic
%ejection events during the early evolution of young stellar objects,
%for each set of 1-mag, 2-mag, 3-mag, and 4-mag bursts in our models
%we calculate durations of quiescent phases between bursts - $\Delta\tau$.
The normalized distribution of $\Delta\tau_{\rm mod}$ in model~1 is presented in Figure \ref{fig:4}
with the filled histograms. More specifically, the upper-left panel presents
the normalized distribution of $\Delta\tau_{\rm mod}$ for bursts of all four magnitudes, 
while the other three panels show the normalized distributions for the bursts of 
increasingly higher magnitudes as indicated in the legends. 
Interestingly, $\Delta\tau_{\rm mod}$ is characterized by a bi-modal distribution 
with one maximum at $\approx 100$~yr and the other maximum at 
$\mathrm{a~few}~\times (10^{3}-10^4)$~yr. The bi-modality is a direct 
consequence of the isolated and clustered burst modes \citep{VB2015}. In the former, 
quiescent periods between luminosity bursts are long (on the order of thousands to tens 
of thousand years), whereas in the latter
bursts occur one after another on time scales of tens to hundred years. 
We note that the bi-modality diminishes for strong 4-mag bursts because
they rarely occur in the clustered mode. A similar bi-modal distribution of 
$\Delta\tau_{\rm mod}$ was also found in model~2.

The solid lines in Figure~\ref{fig:4} present the distribution of
$\Delta\tau_{\rm obs}$ (uncorrected for inclination) for
the jet of CARMA~7. We calculated the
distribution of $\Delta\tau_{\rm obs}$ taking the corresponding data for both the northern and
southern parts of the jet, rather than constructing separate distributions,
to increase statistics and decrease the noise of the distribution. In contrast to
the model $\Delta\tau_{\rm mod}$ distribution, the observed distribution of $\Delta\tau_{\rm obs}$
does not show bi-modality and has a maximum at $\approx300-400$~years, which falls almost 
in between the two peaks in the model distribution. 
The lack of bi-modality can be attributed to relatively short dynamic timescales
of the knots (more distant knots might have dissipated due to interaction with the 
ambient medium), while the mismatch in the maxima 
of the model and observed distributions might be less severe if we applied a correction 
for inclination. 

\begin{table*}
\caption{\label{tab:4}K-S test results for the entire jet (the sum of the northern and southern parts)}
\centering{}%
\begin{tabular}{|c|c|c|c|c|c|c|c|c|}
\hline 
 & \multicolumn{2}{c|}{(1-4)-mag} & \multicolumn{2}{c|}{(2-4)-mag} & \multicolumn{2}{c|}{(3-4)-mag} & \multicolumn{2}{c|}{4-mag}\tabularnewline
\hline 
\hline 
 & P-value & \textit{i} & P-value & \textit{i} & P-value & \textit{i} & P-value & \textit{i}\tabularnewline
\hline 
Model 1 & 0.941 & $65^\circ$ & 0.794 & $75^\circ$ & 0.927 & $70^\circ$ & - & - \tabularnewline
\hline 
Model 2 & 0.674 & $55^\circ$ & 0.934 & $75^\circ$ & 0.929 & $80^\circ$ & 0.963 & $80^\circ$  \tabularnewline
%\hline 
%Model 3 & 0.687 & $25^\circ$ & 0.653 & $70^\circ$ & - & - & - & -\tabularnewline
\hline 
\end{tabular}
\end{table*}

\begin{figure}
\begin{centering}
\includegraphics[width=1\columnwidth]{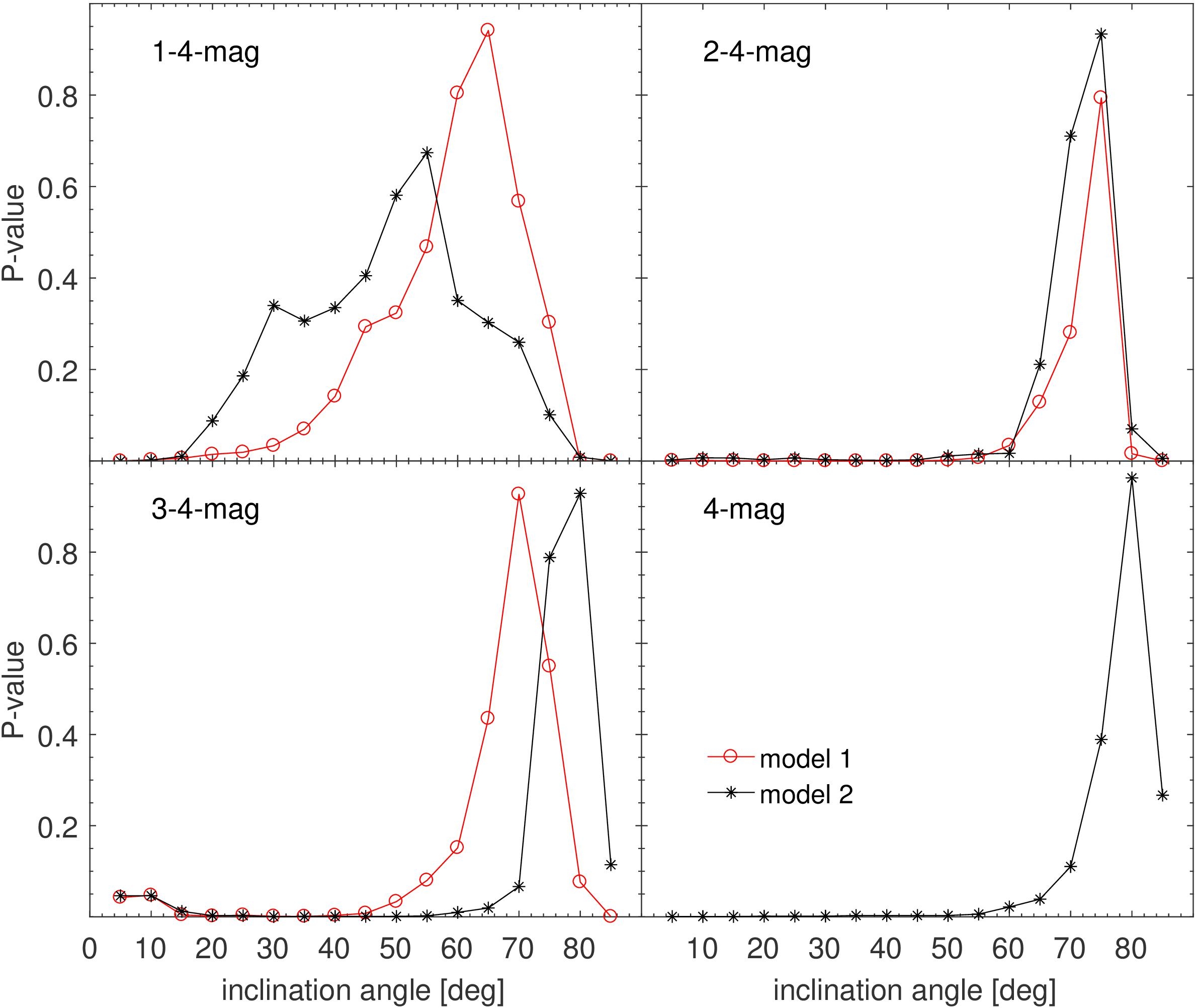} 
\par\end{centering}
\caption{\label{fig:5a}P-values of the K-S test using the unbinned observational and 
theoretical data sets of 
$\Delta\tau_{\rm mod}$ and $\Delta\tau_{\rm obs}$ as a function of (unknown) inclination angle
in CARMA-7. The results for different models are distinguished by lines with different color 
as indicated in the legend. Luminosity bursts comprising various magnitude ranges
(1-4-mag, 2-4-mag, 3-4-mag, and 4-mag) are considered.}
\end{figure}

The inclination angle of the jet with respect to the line of sight in CARMA~7 is poorly constrained.
Therefore, we performed the K-S test using the unbinned observational and theoretical data sets of 
$\Delta\tau_{\rm mod}$ and $\Delta\tau_{\rm obs}$ to find the inclination angle
at which both distributions agree best. For the observational data, we applied an 
inclination angle correction from $i=5^\circ$ to $i=85^\circ$ with a step of $5^\circ$. 
For the model data, we  retained $\Delta\tau_{\rm mod}$ that are equal to or shorter than 
the maximum observed difference in dynamical timescales of the knots 
$\Delta\tau_{\rm obs}^{\rm max}$, the latter also corrected for the corresponding inclination. 
%This truncation of the model data is meant to exclude the bursts with long time spacings, 
%for which we have no observational counterparts.
Here and in the following text, we use the truncated set of model data, because the 
jet structure in CARMA~7 was analyzed only for a narrow field of view. This means that 
more distant knots with longer $\Delta\tau_{\rm obs}$ may exist in CARMA~7, for which we have 
at present no information.
Counting in the model data with $\Delta\tau_{\rm mod}>\Delta\tau_{\rm obs}^{\rm max}$ 
could lead to potentially wrong conclusions.

Figure~\ref{fig:5a} presents the results of the K-S test (P-values) for both considered
models as a function of the inclination angle. We considered bursts
of various magnitudes, but excluded the model data for which the statistics was too poor 
to make firm conclusions (model~1, 4-mag). In all cases, the K-S test
has a clear peak value (that is higher than the minimum value of 0.05 required 
to formally pass the test) at a certain inclination angle. The resulting P-values and best-fit 
inclination angles $i$ for both models are summarized in Table~\ref{tab:4}. 
The best-fit inclination angle increases with the increasing amplitude of the bursts, but
is constrained within a range of $55^\circ-80^\circ$. In fact, if we exclude
the 1-mag bursts, the window of best-fit inclination angles becomes even narrower, 
$75^\circ-80^\circ$.
The wider spread of the inferred inclination angles for the models that include 1-mag
bursts may indicate these bursts are too weak to create a significant response in the jet
and produce notable knots. The inclusion of these weak bursts in the statistical analysis
might have created some sort of a 'noise', which spread inferred inclination angles.
We note that these inclination angles are only rough estimates based on an assumption 
of a causal link between
accretion bursts and jet knots, and need to be confirmed by independent measurements.

\begin{figure}
\begin{centering}
\includegraphics[width=1\columnwidth]{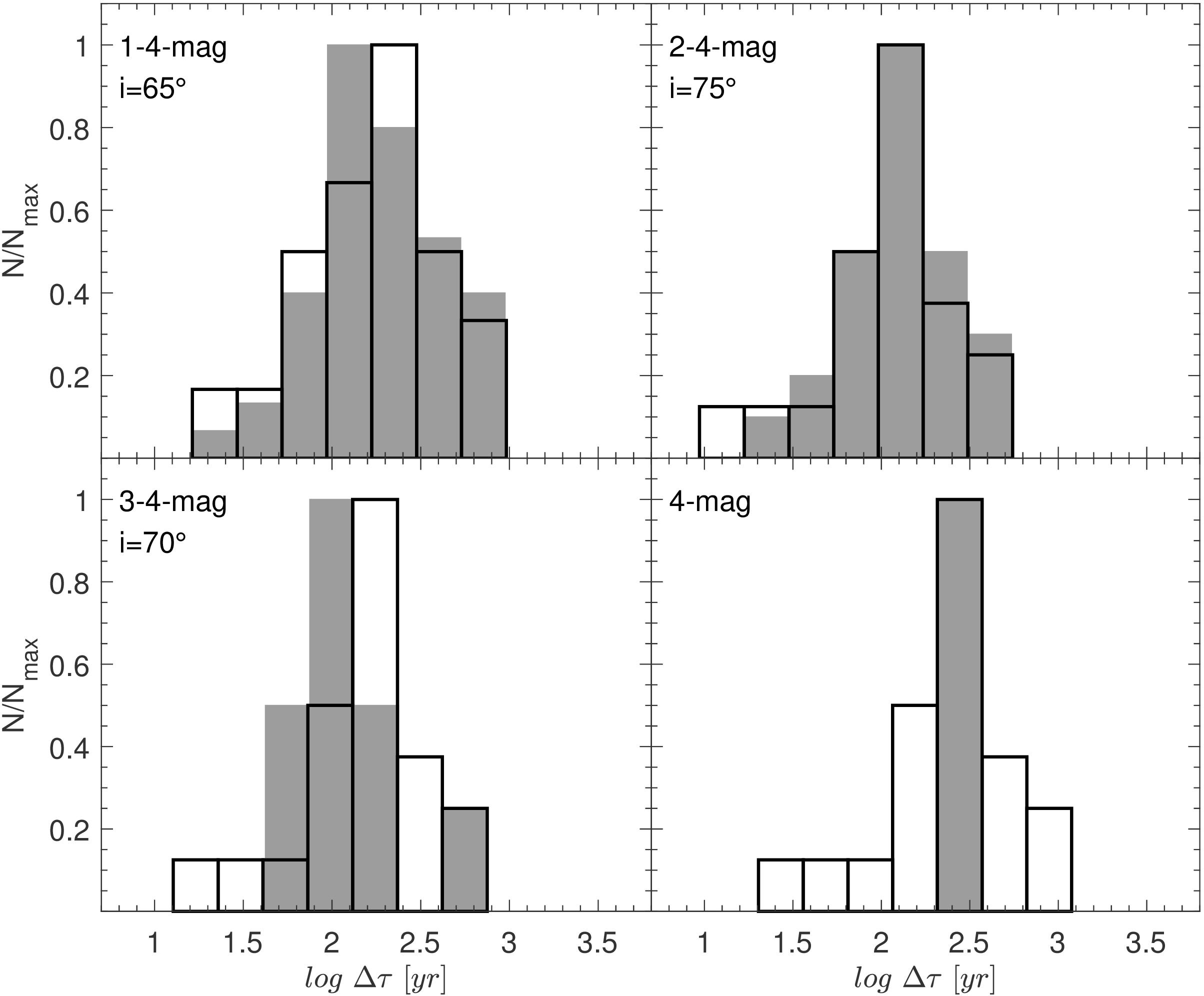} 
\par\end{centering}
\caption{\label{fig:5}Similar to Figure \ref{fig:4}, but with the observational data 
$\Delta\tau_{\mathrm{obs}}$ corrected for the best-fit inclination angles as indicated in Table~\ref{tab:4}.}
\end{figure}

\begin{figure}
\begin{centering}
\includegraphics[width=1\columnwidth]{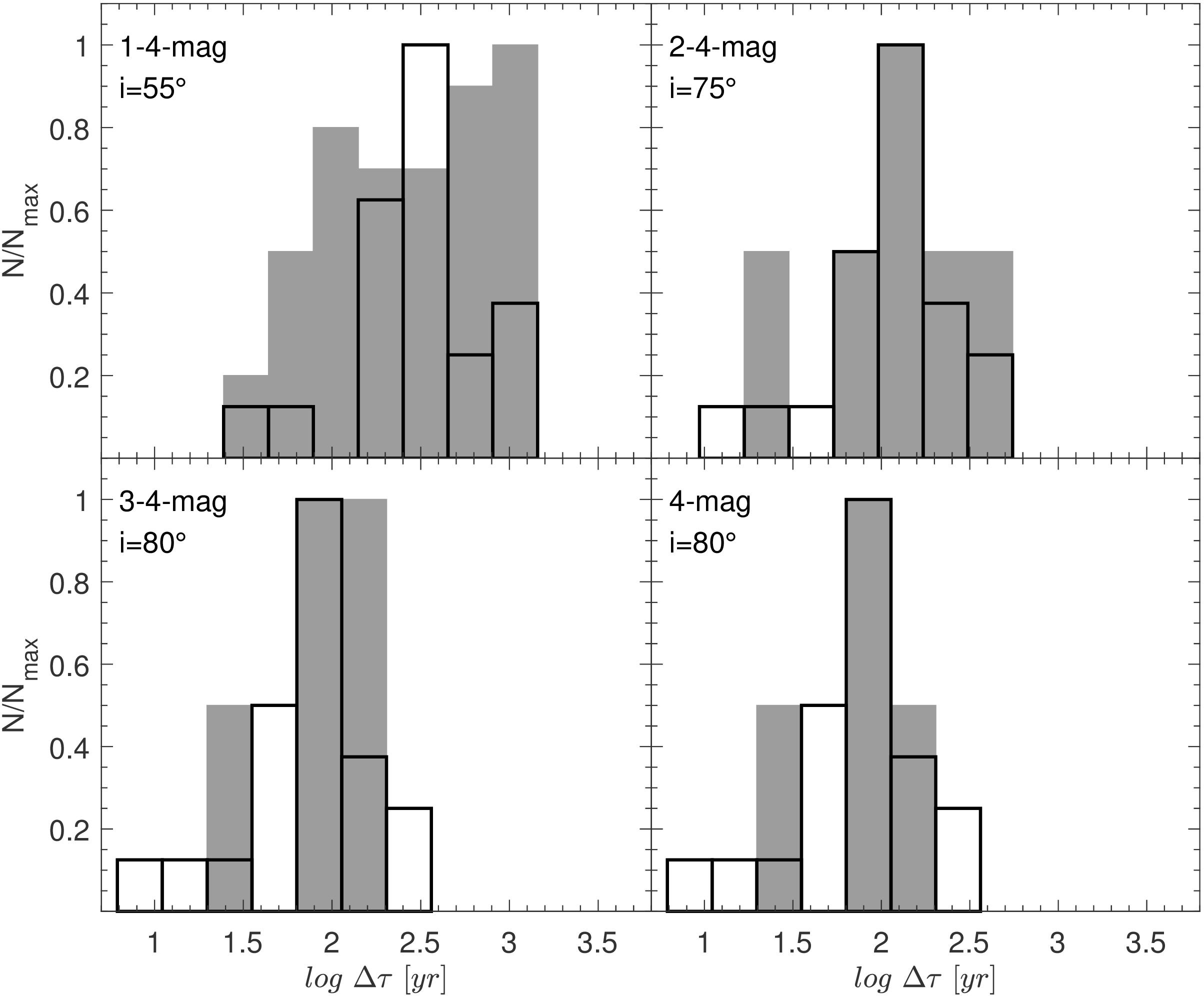} 
\par\end{centering}
\caption{\label{fig:6}Similar to Figure~\ref{fig:5}, but for model~2. }
%The model data are shown only for $\Delta\tau_{\rm mod}\le\Delta\tau_{\rm obs}^{\rm max}=550$~yr.
%The correction of observational data for an inclination angle of $i=75^\circ$ is applied.}
\end{figure}

Figures~\ref{fig:5} and \ref{fig:6} show the model and observational distributions of $\Delta\tau_{\rm mod}$ and $\Delta\tau_{\rm obs}$ after correction for the best-fit inclination angle for models 1 and
2, respectively. The model
data sets were cut, so that $\Delta\tau_{\rm mod}\le\Delta\tau_{\rm obs}^{\rm max}$, where  
$\Delta\tau_{\rm obs}^{\rm max}$ was also corrected for the corresponding inclination.
Clearly, the observational distribution of $\Delta\tau_{\rm obs}$ now fits better to our 
theoretical predictions. We note that remained model data correspond to the clustered 
burst mode with a peak of $\Delta\tau_{\rm mod}$ at a few$\times10^2$~yr.
We conclude that strong luminosity bursts ($>1$-mag) can match the observed
distribution of $\Delta\tau_{\rm obs}$ in the knots of CARMA~7 for inclination angles
that are constrained within a rather narrow range, $i=70^\circ-80^\circ$. The
range of inferred inclination angles becomes somewhat wider ($i=55^\circ-80^\circ$), 
if weaker luminosity bursts (1-mag) are also considered.
The independent knowledge of the inclination angle in CARMA~7 is needed to further constrain the
luminosity burst strengths that can reproduce the observed time spacings between the knots.

Given our constrains on the inclination angle, we can estimate the time passed since the last burst.
The shortest dynamical time (uncorrected for the inclination) for the nearest knot is 100~yr.
For the inclination angles in the $i=55^\circ-80^\circ$ range, the resulting times passed since the
last burst are $\approx17$--70~yr. The magnitude of the last burst is difficult to constrain 
from our analysis. However, if we assume that the last burst was sufficiently strong 
($\ge 3$-mag or more than a factor of 15 in luminosity), then it might have heated notably the surrounding
envelope (CARMA~7 is a class 0~object), thus evaporating certain chemical species, such as CO, 
from dust grains. The freeze-out time of the gas-phase CO back on dust grains 
is much longer, on the order of hundreds
to thousand years \citep{Visser2012,Vorobyov2013,Frimann2017}. 
Therefore, the possible overabundance of the gas-phase CO in the envelope 
of CARMA-7, as compared to what can be expected from the current luminosity, can be used to 
confirm the recent luminosity burst in this object. Even if the latest burst was not sufficiently 
strong to produce a notable heating in the envelope, more distant bursts that occurred no longer than
a few hundred years ago may still be stronger and therefore may leave chemical signatures in the envelope.

%\begin{figure}
%\begin{centering}
%\includegraphics[width=1\columnwidth]{figures/fig7_new.eps} 
%\par\end{centering}
%\caption{\label{fig:7}Similar to Figure~\ref{fig:5}, but for model~3. }
%%The model
%%data are shown only for $\Delta\tau_{\rm mod}\le\Delta\tau_{\rm obs}^{\rm max}=2940$~yr.
%%The correction of observational
%%data for an inclination angle of $i=35^\circ$ is applied}
%\end{figure}

\begin{figure}
\begin{centering}
\includegraphics[width=1\columnwidth]{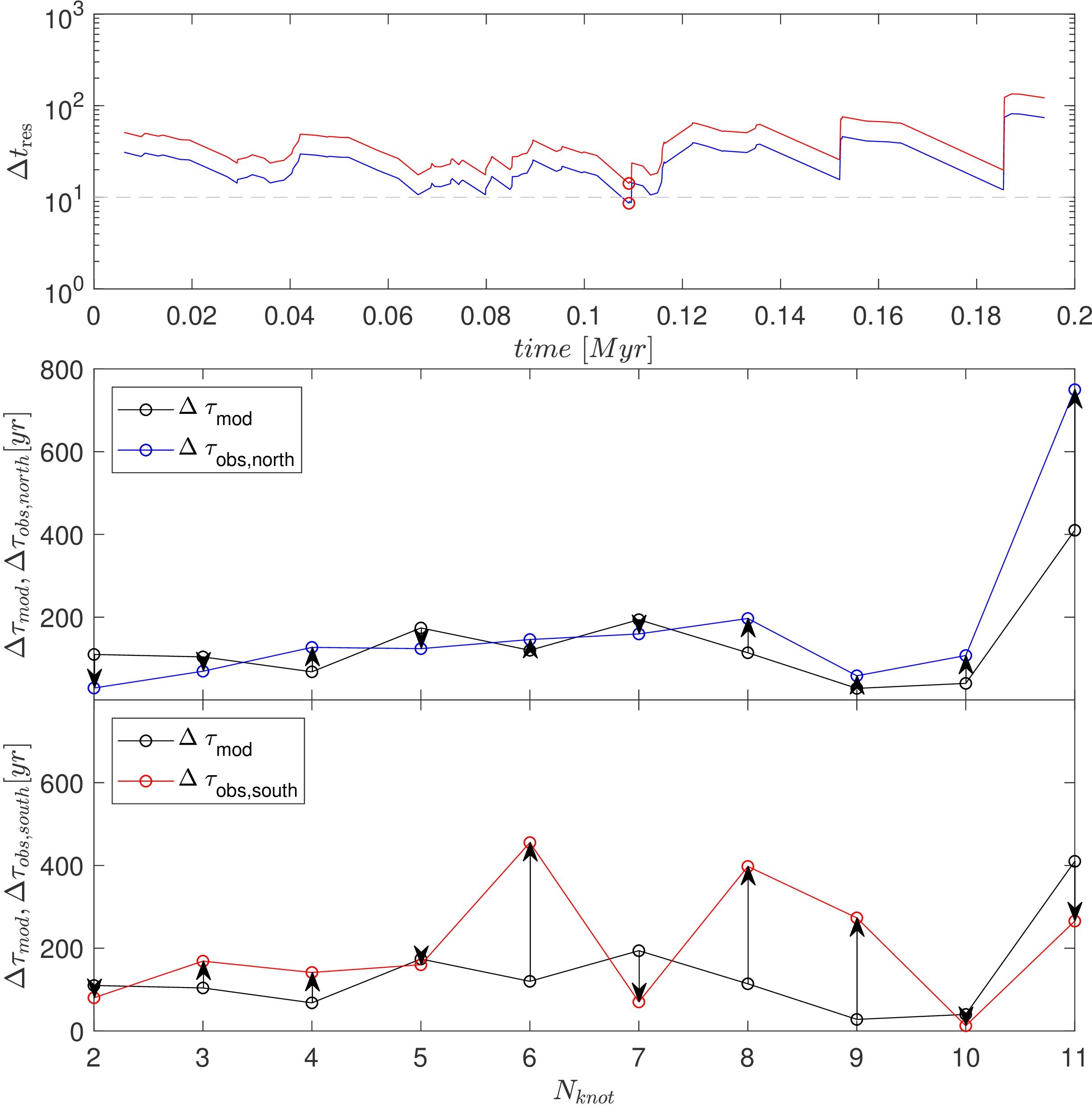} 
\par\end{centering}
\caption{\label{fig:8} Top panel: the residual times left after subtracting the observed sequences 
of $\Delta\tau_{\rm obs}$ from the model sequence of $\Delta\tau_{\rm mod}$ in model~1 
and then normalized to the maximum dynamical timescale of the knots in CARMA~7. The blue 
and red curves
correspond to the northern and southern knots, respectively.
The red circles mark the minimum values which correspond to the best fit between the observational
and model data. Middle and bottom panels: $\Delta\tau_{\rm obs}$
and $\Delta\tau_{\rm mod}$ as a function of the knot serial number at the time instance 
when the best fit between the observed and model data was found. 
The middle panel provides the comparison
for the northern knots, while the bottom panel -- for the southern knots.
The arrows show the mismatch between the individual data pairs. }
\end{figure}

A good agreement between the model and observed distributions of $\Delta\tau$ (for certain inclination
angles) tells us that the time spacings between the bursts 
in our models are similar in magnitude to the differences in dynamical timescales of the observed knots.
However, we still do not know if we can reproduce the exact sequence of
$\Delta\tau_{\mathrm{obs}}$ for 11 northern and 11 southern knots in CARMA~7.
To check this, we first calculate the sum
\begin{equation}
\Delta t_{\rm res} = \sum_{i=1}^{11} { \left| \Delta\tau_{{\rm mod},i} - \Delta\tau_{{\rm obs},i} \right| \over 
\tau_{\rm obs}^{\rm max} },
\end{equation} 
which is the residual time left after subtracting the observed sequences of $\Delta\tau_{\rm obs}$ from
the model sequence of $\Delta\tau_{\rm mod}$ and then normalized
to the maximum dynamical timescale of the knots $\tau_{\rm obs}^{\rm max}$.
Because there are more luminosity bursts in our models than knots in 
CARMA~7, we shifted the observed sequence of $\Delta\tau_{\rm obs}$ along the line of increasing
time in our models to find the time instance when the residual $\Delta t_{\rm res}$ is minimal,
indicating the time instance when the best match between the model and observed sequences of $\Delta\tau$
is achieved. We performed this analysis for both models, but show here only the results 
for model~1, for which the best agreement was found. 
For the model data, we used luminosity bursts of all magnitudes (1-4-mag) 
and for the observational data we used 11 northern and southern knots 
The observational data were corrected for the best-fit
inclination found from the K-S test (see Table~\ref{tab:4}).
The corresponding values of $\Delta t_{\rm res}$ as a function of time (elapsed from the
instance of protostar formation) are shown in 
the top panel of Figure~\ref{fig:8} and the minimum of $\Delta t_{\rm res}$ is marked
by the red circle. There are no data for $t>0.2$~Myr
because there are less than 11 bursts in model~1 left at advanced evolutionary times.

The middle and bottom panels in Figure \ref{fig:8} show $\Delta\tau_{\rm obs}$
and $\Delta\tau_{\rm mod}$ as a function of the knot serial number at the time instance 
when the best fit between
the observed and model data was found (marked with the red circles in the top panel).
All 11 knots in the northern (middle panel) and southern (bottom panel) jet 
were used for comparison with the model data, but only 10 are shown because we calculate 
the difference in 
dynamical timescales between the adjacent knots (i.e., 2-1, 3-2, etc.). 
Clearly, for some knots, the time spacings between the bursts 
$\Delta\tau_{\rm mod}$ differ substantially from the differences in dynamical timescales 
between the knots  $\Delta\tau_{\rm obs}$. For the northern jet, 
the mismatch shown by the arrows is greater than 
$\Delta\tau_{\rm obs}$ itself for knot 2 and it is comparable to $\Delta\tau_{\rm obs}$ for
knots 4, 8, and 11,  while other knots show good agreement 
(e.g., 3, 6, 7, and 9). For the southern jet, the agreement is particularly worse 
for knots 6, 8, and 9. 
The northern-jet knots close to the protostar show a more linear $\Delta\tau_{\rm obs}$  vs. 
$\tau_{\rm obs}$ correlation \citep[see fig. 2e in][]{Plunkett2015}, which matches better the modeling results. The difference in the time spacings between the northern and southern 
jets could be due to different inclinations of the two branches due to precession. 
Similarly, the other model revealed good correlation only for part of the knots in CARMA~7.  
Nevertheless, we emphasize that finding a perfect match
on the basis of only two models is probably a very unlikely event. The stochastic nature of accretion
bursts in gravitationally unstable disks and the effect of environment on the jet/outflow 
propagation can complicate the comparison.   
We will continue searching for a better match once more
models become available.

%\section{Discussion and model caveats}
%\label{discuss}

\section{Conclusions}
\label{summary}
In this paper, we studied numerically the mass accretion history of (sub)-solar
mass protostars during the embedded phase of evolution using numerical hydrodynamics
simulations of gravitationally unstable protostellar disks in the thin-disk limit. 
The accretion variability
resulting from gravitational instability and fragmentation of protostellar disks
was analyzed to obtain the characteristics of accretion and luminosity bursts
that occurred during the embedded phase. In particular, we identified luminosity bursts with
different magnitudes (from 1-mag to $>4$-mag corresponding to an increase in luminosity
from a factor of 2.5 to $>39$) and calculated time spacings between the bursts 
$\Delta\tau_{\rm mod}$.
We further compared $\Delta\tau_{\rm mod}$ with the differences in dynamical timescales of
knots $\Delta\tau_{\rm obs}$ recently detected in the jet of CARMA~7, a young protostar in the 
Serpens South cluster  \citep{Plunkett2015}. More specifically, we compare the model and observed 
distribution functions and also the exact sequences of $\Delta\tau_{\rm mod}$ 
and $\Delta\tau_{\rm obs}$.  We aimed at investigating a possible 
causal link between the episodic mass accretion onto the protostar and the knotty jet structure.
Our results can be summarized as follows.

\begin{itemize}

\item Gravitationally unstable protostellar disks are characterized by time-varying protostellar 
accretion with episodic bursts. Accretion variability is strong in the early evolution, but subsides
with time as the gravitational instability weakens because of diminishing mass infall from the envelope. Artificially turning off disk self-gravity (and hence gravitational instability) results in significant
reduction of accretion variability.

\item The time spacings between the luminosity bursts $\Delta\tau_{\rm mod}$ in gravitationally
unstable disks show a bi-modal distribution, with the first peak at $\approx 100$~yr and the second peak at $\mathrm{a~few}~\times (10^{3}-10^4)$~yr, depending on the strength of the bursts. 
The bi-modality is caused by two modes of luminosity bursts: isolated and clustered ones 
\citep[see][for detail]{VB2015}. In particular, 
the isolated bursts are characterized by long quiescent periods, whereas
clustered bursts occur one after another on time scales of $\mathrm{a~few}~\times (10-10^2)$~yr.

\item The distribution of $\Delta\tau_{\rm mod}$ in our models can be fit reasonably well
to the distribution of differences in dynamical timescales of the knots 
$\Delta\tau_{\rm obs}$ in the protostellar jet of CARMA-7, if a correction for the (yet unknown) 
inclination angle is applied to the observational data set and the model data
are truncated to retain only clustered bursts. The K-S test on the 
unbinned model  and observational data sets suggests a narrow range of the best-fit 
inclination angles 
($i=75^\circ-80^\circ$), if strong luminosity bursts ($>1$-mag) are considered.
The range of inferred inclination angles becomes somewhat wider ($i=55^\circ-80^\circ$), 
if weaker luminosity bursts (1-mag) are also considered.
This may indicate that 1-mag bursts introduce some sort of a 'noise' and are in fact too weak 
to produce notable knots.

% Based on the K-S test comparing the unbinned model and observational
%data sets (the latter corrected for inclination angles from $5^\circ$ to $85^\circ$),
%the best fit is found for the inclination angles of $i=70^\circ, 75^\circ$ and $35^\circ$
%for models 1, 2, and 3, respectively. The K-S test, however, can occasionally fails on some data sets.

\item Notwithstanding a good agreement between the model and observed 
distributions of $\Delta\tau_{\rm mod}$ and $\Delta\tau_{\rm obs}$, the exact sequences
of time spacings between the luminosity bursts in our models and knots in the jet of CARMA~7
were found difficult to match. More models and observational data are needed to 
further explore this issue.

\end{itemize}

Given our constrains on the inclination angle, we estimate the times passed since the
last luminosity burst to be $\approx17$--70~yr. This is much shorter than the typical freeze-out time
of CO in the envelope, on the order of hundreds to thousands years \citep{Visser2012,Vorobyov2013,Rab2017}.
Recent surveys of deeply embedded protostars \citep[e.g.][]{Jorgensen2015,Frimann2017} indicate 
that a notable fraction of sources show extended CO emission that is inconsistent with their 
current luminosity,
implying a recent luminosity burst that heated the envelope and evaporated CO, which is currently in
the process of re-freezing.
This means that we may expect the overabundance of the gas-phase CO in the envelope of CARMA~7
(recall that this is a Class~0 object) if the most recent bursts are strong enough to 
evaporate this species in the envelope. The search for an extended CO emission in CARMA~7 can therefore
confirm the recent luminosity burst in this object.

%Provided that the most recent bursts in CARMA~7 are sufficiently strong to evaporate CO in the envelope,
%the possible overabundance of this gas-phase species as compared to what can be expected from 
%the current luminosity can be used to confirm the recent luminosity burst in this object. 

Finally, we note that we have not taken into account some important effects.
For instance, the luminosity bursts in our models are triggered by disk gravitational
fragmentation with subsequent infall of gaseous clumps on the star. Other
luminosity bursts mechanisms can operate concurrently (and also in disks stable to fragmentation) and increase the number of luminosity bursts. In addition, varying velocity of the blobs ejected 
at different epochs together with the inhomogeneous ambient medium, 
may lead to complex mutual interactions of the blobs, which were not taken into account.
Therefore, deceleration of the knots and also interactions of knots with each
other (collisions) should be considered in later works.
In \citet{Plunkett2015}, a small region near CARMA-7 was only analyzed.  Beyond that area, the 
outflows are confused with other surrounding outflows from nearby protostars.  There might be 
outflow knots with longer time intervals than what was studied in \citet{Plunkett2015}.

\section{Acknowledgments}
 The authors are thankful to the anonymous referee for constructive comments that helped to
improve the manuscript.
This work was supported by the Austrian Science Fund (FWF) under research grant I2549-N27.
OD acknowledges support from the Austrian Research Promotion Agency in the framework of the Austrian Space Application Program (FFG-854025).
The simulations were performed on the Vienna Scientific Cluster (VSC-2
and VSC- 3), on the Shared Hierarchical
Academic Research Computing Network (SHARCNET), and on
the Atlantic Computational Excellence Network (ACEnet).
%V. Elbakyan acknowledges support from the Russian Ministry of Education
%and Science Grant 3.5602.2017.

\label{lastpage} 

\begin{thebibliography}{}


%\bibitem[\protect\citeauthoryear{Plunkett et al.}{2015}]{Plunkett2015}
%Plunkett, A. L., Arce, H. G., Mardones, D., van Dokkum, P., Dunham, M. M., et al. 2015, Nature, 527,
%70

\bibitem[\protect\citeauthoryear{Andr\'e et al.}{1993}]{Andre1993}
Andr\'e, P., Ward-Thompson, D., \& Barsony, M. 1993, ApJ, 406, 122

\bibitem[\protect\citeauthoryear{Arce et al.}{2007}]{Arce2007}
Arce, H. G., Shepherd, D., Gueth, F., Lee, C.-F., Bachiller, R., Rosen, A., \& 
Beuther, H. 2007, in Protostars and Planets V, eds: B. Reipurth, D. Jewitt, and K. Keil, 
University of Arizona Press, Tucson, 951, 245

\bibitem[\protect\citeauthoryear{Armitage}{2001}]{Armitage2001}
Armitage, P. J., Livio, M., \& Pringle, J. E. 2001, MNRAS, 324, 705

\bibitem[\protect\citeauthoryear{Armitage}{2016}]{Armitage2016}
Armitage, P. J. 2016, ApJ, 833, 15

\bibitem[\protect\citeauthoryear{Audard et al.}{2014}]{Audard2014}
Audard, M., Ábrahám, P., Dunham, M. M., et al. 2014, in Protostars and
Planets VI, ed. H. Beuther, R. S. Klessen, C. P. Dullemond, \& T. Henning
(Tucson, AZ: Univ. Arizona Press), 387

%\bibitem[\protect\citeauthoryear{Baraffe \& Chabrier}{2010}]{Baraffe2010}
%Baraffe, I., \& Chabrier, G. 2010, A\&A, 521, 44

\bibitem[\protect\citeauthoryear{Bell \& Lin}{1994}]{BL1994}
Bell, K. R., \& Lin, D. N. C. 1994, ApJ, 427, 987

\bibitem[\protect\citeauthoryear{Bonito et al.}{2010}]{Bonito2010}
Bonito, R., Orlando, S., Peres, G., Eisl¨offel, J., Miceli, M., \& Favata, F. 2010, A\&A, 511, 42

\bibitem[\protect\citeauthoryear{Caselli et al.}{2002}]{Caselli2002}
Caselli, P., Benson, P. J., Myers, P. C., \& Tafalla, M. 2002, ApJ, 572, 238

\bibitem[\protect\citeauthoryear{Cieza et al.}{2017}]{Cieza2017}
Cieza, L. A., Ru\'iz-Rodr\'iguez D., Perez, S., et al. 2017, arXiv:1711.08693

\bibitem[\protect\citeauthoryear{D'Angelo et al.}{2012}]{Dangelo2012}
D'Angelo C. R. and Spruit H. C. 2012, MNRAS,
420, 416.

\bibitem[\protect\citeauthoryear{Dapp \& Basu}{2009}]{Dapp09}
Dapp, W. B., \& Basu, S. 2009, MNRAS, 395, 1092

\bibitem[\protect\citeauthoryear{Dopita}{1978}]{Dopita1978}
Dopita, M. A. 1978, A\&A, 63, 237

\bibitem[\protect\citeauthoryear{Dunham \& Vorobyov}{2012}]{Dunham2012}
Dunham, M. M., \& Vorobyov, E. I. 2012, ApJ, 747, 52

\bibitem[\protect\citeauthoryear{Elbakyan et al.}{2016}]{Elbakyan2016}
Elbakyan, V. G., Vorobyov, E. I., Glebova, G. M 2016, Astronomy Rep., 60, 679

\bibitem[\protect\citeauthoryear{Ellerbroek et al.}{2014}]{Ellerbroek2014}
Ellerbroek, L. E., Podio, L., Dougados, C., Cabrit, S., Sitko, M. L., et al. 2014, 563, 87

\bibitem[\protect\citeauthoryear{Enoch et al.}{2009}]{Enoch2009}
Enoch, M. L., Evans, N. J., II, Sargent, A. I., \& Glenn, J. 2009, ApJ, 692,
973

\bibitem[\protect\citeauthoryear{Evans et al.}{2009}]{Evans2009}
Evans, N. J., II, Dunham, M. M., Jørgensen, J. K., et al. 2009, ApJS, 181, 321


\bibitem[\protect\citeauthoryear{Frimann et al.}{2017}]{Frimann2017}
Frimann, S, Jorgensen, J. K., Dunham, Michael M., et al. 2017, A\&A 602, 120

\bibitem[\protect\citeauthoryear{Frank et al.}{2014}]{Frank2014}
Frank, A., Ray, T. P., Cabrit, S., Hartigan, P., Arce, H. G., et al. 2014,  
in Protostars and Planets VI, eds. H. Beuther, R. S. Klessen, C. P. Dullemond, and T. Henning, 
University of Arizona Press, Tucson, 914, 451


\bibitem[\protect\citeauthoryear{Goodson et al.}{1999}]{Goodson1999}
Goodson, A. P., Bohm, K.-H., Winglee, R. M. 1999, ApJ, 524, 142

\bibitem[\protect\citeauthoryear{Herzeg et al.}{2017}]{Herczeg2017}
Herczeg, G. J., Johnstone, D. I., Mairs, S. et al. 2017, ApJ, submitted


\bibitem[\protect\citeauthoryear{Hsieh et al.}{2016}]{Belloche2016}
Hsieh, T.-H., Lai, S.-P., Belloche, A., Wyrowski, F. 2016, ApJ 826, 68

\bibitem[\protect\citeauthoryear{Kenyon et al.}{1990}]{Kenyon1990}
Kenyon, S. J., Hartmann, L.W., Strom, K. M., \& Strom, S. E. 1990, ApJ, 99, 869

\bibitem[\protect\citeauthoryear{K\'osp\'al et al.}{2017}]{Kospal2017}
K\'osp\'al, A., et al. 2017, ApJ, 843, 45

\bibitem[\protect\citeauthoryear{Kratter et al.}{2008}]{Kratter2008}
Kratter, K. M., Matzner, C. D., \& Krumholz, M. R. 2008, ApJ, 681, 375

\bibitem[\protect\citeauthoryear{Kraus et al.}{2016}]{Kraus2016}
Kraus, S., Caratti o Garatti, A., Garcia-Lopez, R., Kreplin, A., Aarnio, A. et al. 2016, MNRAS, 462, 61

\bibitem[\protect\citeauthoryear{J{\o}rgensen et al.}{2015}]{Jorgensen2015}
J{\o}rgensen, J. K., Visser, R., Williams, J. P., \& Bergin, E. A. 2015, A\&A, 579,
A23

\bibitem[\protect\citeauthoryear{Larson}{1969}]{Larson1969}
Larson, R. 1969, MNRAS, 145, 271

\bibitem[\protect\citeauthoryear{Lee}{2007}]{Lee2007}
Lee, J.-E. 2007, J. Korean Astron. Soc., 40, 83

\bibitem[\protect\citeauthoryear{Meyer et al.}{2017}]{Meyer2017}
Meyer, D. M.-A., Vorobyov, E. I., Kuiper, R., Kley, W. 2017, MNRAS, 464, 90

\bibitem[\protect\citeauthoryear{Nayakshin \& Lodato}{2012}]{NL2012}
Nayakshin, S., \& Lodato, G. 2012, MNRAS, 426, 70

\bibitem[\protect\citeauthoryear{Pena}{2017}]{Pena2017}
Pena, C. C., Lucas, P. W., Minniti, D. et al. 2017, MNRAS, 465, 3011

\bibitem[\protect\citeauthoryear{Plunkett et al.}{2015}]{Plunkett2015}
 Plunkett, A. L., Arce, H. G., Mardones, D. et al. 2015, Nature, 527, 70 

\bibitem[\protect\citeauthoryear{Ohtani et al.}{2014}]{Ohtani2014}
Ohtani, T., Kimura, S. S., Tsuribe, T., \& Vorobyov, E. I. 2014, PASJ,
66, 1120

\bibitem[\protect\citeauthoryear{Quanz et al.}{2007}]{Quanz2007}
Quanz S. P. et al. 2007, ApJ, 668, 359

\bibitem[\protect\citeauthoryear{Rab et al.}{2017}]{Rab2017}
Rab, Ch., Elbakyan, V., Vorobyov, E., et al. 2017, A\&A, 604, 15

\bibitem[\protect\citeauthoryear{Reipurth \& Cernicharo}{1995}]{Reipurth1995}
Reipurth, B., \& Cernicharo, J. 1995, RMxAA, 1, 43

\bibitem[\protect\citeauthoryear{Reipurth \& Aspin}{1997}]{Reipurth1997}
Reipurth, B., \& Aspin, C. 1997, AJ, 114, 2700 

\bibitem[\protect\citeauthoryear{Safron et al.}{2015}]{Safron2015}
Safron, E. J., Fischer, W. J., Megeath, S. et al. 2015, ApJL, 800, 5


\bibitem[\protect\citeauthoryear{Shu}{1977}]{Shu1977}
Shu, F. H. 1977, ApJ, 214, 488

\bibitem[\protect\citeauthoryear{Stone \& Norman}{1992}]{SN1992}
Stone, J. M., \& Norman, M. L. 1992, ApJS, 80, 753

\bibitem[\protect\citeauthoryear{Visser \& Bergin}{2012}]{Visser2012}
Visser, R., \& Bergin, E. A. 2012, ApJ, 754, 18

\bibitem[\protect\citeauthoryear{Vorobyov \& Basu}{2005}]{VB2005}
Vorobyov, E. I., \& Basu, S. 2005, MNRAS, 360, 675

\bibitem[\protect\citeauthoryear{Vorobyov}{2009}]{Vorobyov2009}
Vorobyov, E. I. 2009, ApJ, 704, 715

\bibitem[\protect\citeauthoryear{Vorobyov \& Basu}{2009}]{VB2009}
Vorobyov, E. I., \& Basu, S. 2009, MNRAS, 393, 822

%\bibitem[\protect\citeauthoryear{Vorobyov}{2010}]{Vorobyov2010}
%Vorobyov, E. I. 2010, ApJ, 713, 1053


\bibitem[\protect\citeauthoryear{Vorobyov \& Basu}{2010}]{VB2010}
Vorobyov, E. I., \& Basu, S. 2010, ApJ, 719, 1896

\bibitem[\protect\citeauthoryear{Vorobyov}{2011}]{Vorobyov2011}
Vorobyov, E. I. 2011, ApJ, 729, 146


\bibitem[\protect\citeauthoryear{Vorobyov et al.}{2013}]{Vorobyov2013}
Vorobyov, E. I., Baraffe, I., Harries, T., \& Chabrier, G. 2013, A\&A, 557, 35 

\bibitem[\protect\citeauthoryear{Vorobyov et al.}{2013}]{Vor2013}
Vorobyov, E. I., Zakhozhay, O. V., \& Dunham, M. M. 2013, MNRAS, 433, 3256

\bibitem[\protect\citeauthoryear{Vorobyov}{2013}]{Vorobyov2013}
Vorobyov, E. I. 2013, A\&A, 552, 129

\bibitem[\protect\citeauthoryear{Vorobyov \& Basu}{2015}]{VB2015}
Vorobyov, E. I., \& Basu, S. 2015, ApJ, 805, 115

\bibitem[\protect\citeauthoryear{Cleveland \& Devlin}{1998}]{Cleveland1998}
Cleveland, W. S., \&  Devlin, S. J. 1998, Journal of the
American Statistical Association, 83, 596

\bibitem[\protect\citeauthoryear{Zhu et al.}{2009}]{Zhu2009}
Zhu, Z., Hartmann, L., \& Gammie, C. 2009, ApJ, 694, 1045

\end{thebibliography}
\end{document}